\documentclass[journal,twoside,web]{ieeecolor}
\usepackage{generic}
\usepackage[utf8]{inputenc}
\usepackage[T1]{fontenc} 
\usepackage{url}

\usepackage{xcolor}
\usepackage[colorlinks]{hyperref}
\hypersetup{
    linkcolor=blue,
    filecolor=magenta, pdfproducer={} 
    urlcolor=teal,
    citecolor=magenta
    }

\usepackage{adjustbox}
\usepackage{subfigure}

\usepackage{comment}
\usepackage{multirow}
\usepackage{pgfplots}
\usepackage{amssymb}
\usepackage{array}
\usepackage{multirow}
\usepackage{cite}
\usepackage{amsmath,amssymb,amsfonts}
\usepackage{algorithmic}
\usepackage{graphicx}
\usepackage{textcomp}
\usepackage{xcolor}

\newcolumntype{L}[1]{>{\raggedright\let\newline\\\arraybackslash\hspace{0pt}}m{#1}}
\newcolumntype{C}[1]{>{\centering\let\newline\\\arraybackslash\hspace{0pt}}m{#1}}
\newcolumntype{R}[1]{>{\raggedleft\let\newline\\\arraybackslash\hspace{0pt}}m{#1}}

\usepackage{float}
\usepackage[linesnumbered,ruled]{algorithm2e}
\pgfplotsset{compat=1.12}
\usetikzlibrary{pgfplots.statistics}
\usetikzlibrary{positioning,chains}

\usepackage{pgfplots}
    \usepgfplotslibrary{statistics}
    \pgfplotsset{compat=1.8}
    \pgfmathdeclarefunction{fpumod}{2}{%
        \pgfmathfloatdivide{#1}{#2}%
        \pgfmathfloatint{\pgfmathresult}%
        \pgfmathfloatmultiply{\pgfmathresult}{#2}%
        \pgfmathfloatsubtract{#1}{\pgfmathresult}%
        \pgfmathfloatifapproxequalrel{\pgfmathresult}{#2}{\def\pgfmathresult{5}}{}%
    }
    
 \pgfplotsset{width=8cm,compat=1.7}
 \pgfplotsset{legend image post style={sharp plot}}
 \pgfplotsset{every axis legend/.append style={
    at={(1,-0.1)},
    anchor=north east}}
\newcommand{\commentFR}[2][FR]{{\slshape\color{blue}[{\bf #1}: #2]}}

\makeatletter
\def\fnum@figure{\textcolor{subsectioncolor}{\sf Fig.~\thefigure}}
\def\fnum@table{\textcolor{subsectioncolor}{\sf TABLE~\thetable}}
\makeatother

\begin{document}
\bstctlcite{IEEEexample:BSTcontrol}

\title{The CirCor DigiScope Dataset: From Murmur Detection to Murmur Classification}%


\author{Jorge~Oliveira$^\star$, Francesco~Renna, Paulo~Dias~Costa, Marcelo~Nogueira, Cristina~Oliveira, Carlos~Ferreira, Alípio~Jorge,  Sandra~Mattos, Thamine~Hatem, Thiago~Tavares, Andoni~Elola, Ali~Bahrami~Rad, Reza~Sameni,  Gari~D~Clifford, and Miguel~T.~Coimbra%
\thanks{Manuscript received June XX, 2021; revised November YY, 2021. \textit{Asterisk indicates corresponding author.}}%
\thanks{
J. Oliveira is with Universidade Portucalense Infante D. Henrique, Rua Dr. Antonio Bernardino de Almeida, 541 4200-072 Porto, Portugal (email: joliveira@upt.pt). F. Renna is with the Instituto de Telecomunica\c{c}\~{o}es, Faculdade de Ci\^{e}ncias da Universidade do Porto, Rua do Campo Alegre 1021/1055, 4169-007 Porto, Portugal. 
M. Nogueira, C. Ferreira, A. Jorge and M. Coimbra are with INESC TEC, Rua Dr. Roberto Frias, 4200-465 Porto, Portugal. M. Nogueira, A. Jorge, M. Coimbra and P. D. Costa are with Faculdade de Ci\^{e}ncias da Universidade do Porto. 
A.~Elola, A.~Bahrami~Rad, R.~Sameni and G.D.~Clifford are with the Department of Biomedical Informatics, Emory University School of Medicine. A.~Elola is also with the Department of Mathematics in the University of the Basque Country. G.D. Clifford is also with the Department of Biomedical Engineering, Georgia Institute of Technology and Emory University, Atlanta, GA. 
S. Mattos, T. Hatem and T. Tavares are with Círculo do Coração de Pernambuco. 
The corresponding author's email: \url{joliveira@upt.pt}
}
}

\markboth{Preprint. To appear in 
IEEE J Biomed Health Inform. Dec 2021; doi: 10.1109/JBHI.2021.3137048}{J.~Oliveira et al.: The CirCor DigiScope Dataset}

\maketitle

\begin{abstract}
Cardiac auscultation is one of the most cost-effective techniques used to detect and identify many heart conditions. Computer-assisted decision systems based on auscultation can support physicians in their decisions. Unfortunately, the application of such systems in clinical trials is still minimal since most of them only aim to detect the presence of extra or abnormal waves in the phonocardiogram signal, i.e., only a binary ground truth variable (normal vs abnormal) is provided. This is mainly due to the lack of large publicly available datasets, where a more detailed description of such abnormal waves (e.g., cardiac murmurs) exists.  

To pave the way to more effective research on healthcare recommendation systems based on auscultation, our team has prepared the currently largest pediatric heart sound dataset. A total of 5282 recordings have been collected from the four main auscultation locations of 1568 patients, in the process, 215780 heart sounds have been manually annotated. Furthermore, and for the first time, each cardiac murmur has been manually annotated by an expert annotator according to its timing, shape, pitch, grading, and quality. 
In addition, the auscultation locations where the murmur is present were identified as well as the auscultation location where the murmur is detected more intensively. Such detailed description for a relatively large number of heart sounds may pave the way for new machine learning algorithms with a real-world application for the detection and analysis of murmur waves for diagnostic purposes.
\end{abstract}

\section{Introduction}
Cardiovascular disease (CVD) is an umbrella term used to define a heterogeneous group of disorders of the heart and vessels, such as coronary artery disease (CAD), valvular heart disease (VHD), or congenital heart disease (CHD)~\cite{Douglas:2015}. As a whole, CVD is the major cause of death worldwide, accounting for 31\% of all deaths globally~\cite{WHO:2017}. In addition to mortality, CVD severely increases morbidity and causes lifelong disabilities, decreasing the quality of life and potently increases the frequency of hospital admissions, ultimately increasing the economic burden of CVD in healthcare systems and populations \cite{Gheorghe:2018}. 

The majority of the populations in under-developed and developing countries do not have access to an integrated primary healthcare system, as a result diagnosis and treatment of CVD can be delayed, thus potentially leading to early deaths. Furthermore and according to ~\cite{Murphy:2020}, CVD contribute to impoverishment due to health-related expenses and high out-of-pocket expenditure.  This imposes an additional burden on the economies of low-to-middle-income countries. Although CVD incidences in the United States and Europe are high \cite{Benjamin:2019,Wilkins:2017}, over 75\% of CVD-related deaths are estimated to occur in low- and middle-income countries, suggesting a strong link with the socio-economic status \cite{Leal:2006}. In contrast to developed countries, where CAD is more frequent. The CHD and VHD (mainly of rheumatic etiology) are more prevalent in developing countries, primarily due to the lack of prenatal screening programs and access to healthcare. Over 68\,million cases of rheumatic heart disease (RHD) are reported every year, resulting in 1.4\,million deaths. RHD is also the single largest cause of hospital admissions for children and young adults, with observations showing a high prevalence of the disease in young adults \cite{Soler:2000}. 

Despite the development of advanced cardiac monitoring and imaging schemes, cardiac auscultation remains an important first-line screening and cost-effective tool. Cardiac auscultation provides insights of the mechanical activity of the heart. These activities generate sounds that are recorded and saved in the phonocardiogram (PCG) signal. 
Furthermore, cardiac auscultation is a key screening exam \cite{Wealth:2011} and when used properly, it can speed up treatment/referral, thus improving the patient's outcome and life quality \cite{MANGIONE_2001}. However, auscultation is a difficult skill to master, requiring long and hard training accompanied by continuous clinical experience. Moreover, the emergence of new imaging techniques and the significant decrease of valve diseases in developed countries reduced the clinical application of cardiac auscultation~\cite{Narula:2018}. As a result, auscultation has been recently neglected, and a great reduction of this clinical skill has been observed over the last decades~\cite{MANGIONE_2001}. In underprivileged countries, the scenario is similar, with a lack of trained professionals with these skills and the resources to train them. This motivates the use of computer-aided decision systems based on auscultation to accelerate the timely diagnosis and referral of patients to specialized treatment centers~\cite{Falleni:2017}. 
 
A fundamental step in developing computer-aided decision systems for CVD screening from cardiac auscultation consists in collecting large annotated datasets of heart sounds that can properly represent and characterize murmurs and anomalies in patients with different CVDs. The need for such datasets is specifically crucial when considering the design of modern data-hungry machine learning techniques.

Currently available annotated public PCG datasets have limited scope such that they either provide limited information regarding a general evaluation of the heart sound (normal vs. abnormal) \cite{Clifford:2016}, or the presence/absence of abnormal sounds (e.g., murmurs, clicks, extra sounds)~\cite{Pascal:2013}. Few attempts have been devoted to providing large public datasets with richer characterizations of heart sounds (see e.g., \cite{Xiao:2019}). However, in these cases, heart sounds are only classified based on the underlying cardiac conditions without providing a full characterization of the specific sound signatures encountered in each recording.

The hereby presented dataset aims to address these limitations, thus providing the necessary means to develop novel computer-assisted decision systems, which are able to offer a rich characterization of heart sound anomalies. This is accomplished by collecting a large set of sounds and characterizing them using the same scales and parameters commonly used in clinical practice. The presented dataset was collected during two independent cardiac screening campaigns in the Pernambuco state, Brazil. These campaigns were organized to screen a large pediatric population in Northeast Brazil and to provide the basis for the further development of a telemedicine network \cite{Mattos:2015,Mattos:2018}.





The gathered dataset includes the characterization of each murmur in the recorded heart sounds from different perspectives, including timing, pitch, grading, shape and quality. Moreover, multi-auscultation location recordings are provided for each patient, alongside with the corresponding annotations, which are helpful for inferring the specific source of each murmur \cite{Oliveira:2022}.

The remainder of this paper is organized as follows. In Section ~\ref{par:background}, a background on cardiac auscultation is provided. In Section~\ref{par:datasets}, a brief review of publicly available heart sound datasets is provided. In Sections~\ref{par:proposed} and \ref{par:methods}, methods and the data annotation process are presented. Section~\ref{par:discussion} is devoted to a discussion on the obtained results. Concluding remarks and future perspectives are drawn in Section \ref{par:conclusions}.

\section{Background}
\label{par:background}

\subsection{Cardiac Auscultation}
\begin{figure}[tb]
    \centering
    \includegraphics[width=\columnwidth]{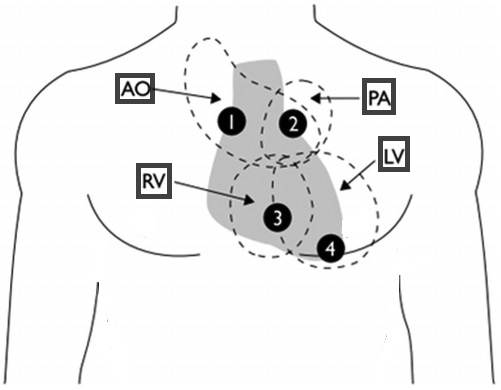}
    \caption{Cardiac auscultation spots (image adapted from \cite{Karnath:2002}); AO = aortic area; LV = left ventricle; PA = pulmonary area; RV = right ventricle; 1 = right second intercostal space; 2 = left second intercostal space; 3 = midleft sternal border (tricuspid); 4 = fifth intercostal space, midclavicular line.}
    \label{cardiacauscultationpoints}
\end{figure}

Normal heart sounds are primarily generated from the vibrations of cardiac valves as they open and close during each cardiac cycle and the turbulence of the blood into the arteries. The anatomical position of heart valves relative to the chest wall defines the optimal auscultation position; as such, a stethoscope should be placed at the following positions \cite{Dornbush:2020}, for auscultation purposes (as illustrated in Figure~\ref{cardiacauscultationpoints}):
\begin{itemize}
    \item \textit{Aortic valve (1)}: second intercostal space, right sternal border;
    \item \textit{Pulmonary valve (2)}: second intercostal space, left sternal border;
    \item \textit{Tricuspid valve (3)}: left lower sternal border;
    \item \textit{Mitral valve (4)}: fifth intercostal space, midclavicular line (cardiac apex).
\end{itemize}

\begin{figure}[t]
  \centering
  \includegraphics[width=0.50\textwidth]{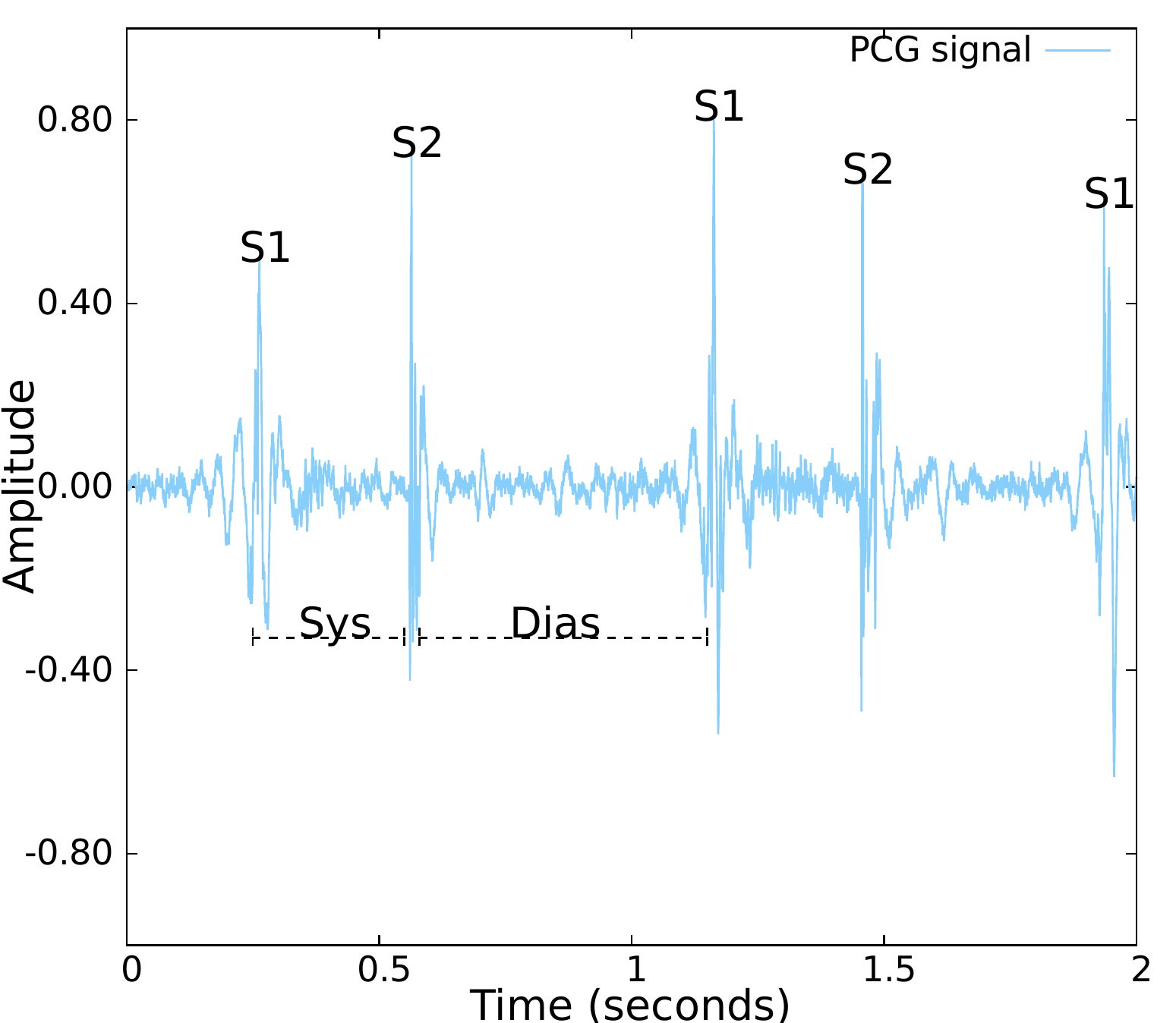}
  \caption{An example of a normalized heart sound recording, the position of the fundamental heart sounds S1 and S2 are displayed and identified. Furthermore, the Systolic (Sys) and the Diastolic (Dias) periods are also displayed and identified.}
  \label{fig:sound_example}
\end{figure}

Blood flowing through these structures creates audible sounds, which are more significant when the flow is more turbulent  \cite{Douglas:2015}.
The first heart sound (S1) is produced by vibrations of the mitral and tricuspid valves as they close in at the beginning of the systole. S1 is audible on the chest wall, and formed by the mitral and tricuspid components ~\cite{Chizner:2008}. Although the mitral component of S1 is louder and occurs earlier, under physiological resting conditions, both components occur closely enough, making it hard to distinguish between them \cite{Dornbush:2020}, an illustration of a S1 sound is provided in Figure \ref{fig:sound_example}.
The second heart sound (S2) is produced by the closure of the aortic and pulmonary valve at the beginning of the diastole. S2 is also formed by two components, with the aortic component being louder and occurring earlier than the pulmonary component (since the pressure in the aorta is higher than in the pulmonary artery). In contrast, unlike S1, under normal conditions the closure sound of the aortic and pulmonary valves can be discernible, due to an increase in venous return during inspiration which slightly delays the pressure increase in the pulmonary artery and consequently the pulmonary valve closure \cite{Dornbush:2020,PRAKASH:1978}, an illustration of a S2 sound is provided in Figure \ref{fig:sound_example}.

\subsection{Clinical Applications}
\label{sec:clinical}
Apart from S1 and S2, additional heart sounds resulting from a turbulent blood flow and rapid movements (closures, openings and resonance) of cardiac structures, and are often associated with murmurs, ``clicks'' and ``snaps'' \cite{Thomas:2020}. Most cardiac murmurs are closely associated with specific diseases such as septal defects, failure of the ductus arteriosus closure in newborns, or defective cardiac valves \cite{Hoeting:2017}, \cite{Riknagel:2017}. These extra sounds may be associated to physiological or pathological conditions, depending on their timing in the cardiac cycle, intensity, shape, pitch, auscultation location, radiation, rhythm, and response to physical exam manoeuvres \cite{Dornbush:2020}. Auscultation is crucial in identifying and distinguishing these features. The ability to accurately describe a murmur can determine whether or not to refer a subject to a cardiologist by identifying specific patterns, isolated or in association with cardiopathies \cite{Thomas:2020}. In the next paragraph, the murmur description of common valve diseases will be described.

\subsubsection{Aortic stenosis}
Aortic stenosis is a narrowing of the aortic valve and often as a result of valve calcification. Furthermore, aging, chronic RHD or a congenital bicuspid aortic valve are some risk factors. During auscultation, Aortic stenosis generates a harsh crescendo-decrescendo systolic murmur heard best at the right upper sternal border (RUSB), with murmur radiating to the carotid arteries \cite{Thomas:2020}.

\subsubsection{Aortic regurgitation}
Aortic regurgitation results from an inefficient Aortic valve closure, allowing the blood to flow back into the right ventricle in a turbulent manner, usually due to aortic root dilation, bicuspid aortic valve, and calcified leaflets. During auscultation, Aortic regurgitation generates a decrescendo-blowing diastolic murmur, heard best at the left lower sternal border (LLSB) \cite{Thomas:2020}.

\subsubsection{Mitral stenosis}
Mitral stenosis is a narrowing of the mitral valve, making blood passing from the left atrium to the left ventricle more difficult. It is often due to chronic RHD and infective endocarditis. During auscultation, Mitral stenosis generates a diastolic murmur, best heard at the apex \cite{Thomas:2020}.

\subsubsection{Mitral regurgitation}
Mitral regurgitation allows the blood to flow back to the left atrium. This is mainly due to defective closure of the leaflets or rupture of the chordae tendineae in patients with infective endocarditis, chronic RHD disease, degenerative valve disease, or myocardial infarction. In auscultation, Mitral regurgitation results in a systolic murmur, best heard at the apex, with radiation to the left axilla \cite{Thomas:2020}.

\subsubsection{Mitral valve prolapse}
Mitral valve prolapse is characterized by a bulge of the leaflets into the left atrium, stopping the valve to close evenly. It can be explained by idiopathic myxomatous valve degeneration, RHD, endocarditis, and Ebstein's anomaly. In auscultation, Mitral valve prolapse causes an early systolic click heard best at the apex, it is often followed by a late systolic murmur \cite{Thomas:2020}.

\subsubsection{Pulmonary stenosis}
Pulmonary stenosis is a narrowing in the pulmonary valve. It is often present in patients with Tetralogy of Fallot, although it may also be present in patients with chronic RHD, congenital rubella syndrome, or Noonan syndrome. Its auscultation is described as a crescendo-decrescendo systolic ejection murmur, heard loudest at the upper left sternal border (LUSB) \cite{Thomas:2020}.

\subsubsection{Tricuspid stenosis}
The tricuspid stenosis is a narrowing in the Triscupid valve. It is often due to intravenous drugs used in infective endocarditis and in carcinoid syndrome treatments. In auscultation, Tricuspid stenosis results in a diastolic murmur, best heard at the LLSB \cite{Thomas:2020}.

\subsubsection{Tricuspid regurgitation}
Tricuspid regurgitation is usually caused by vegetative growth under the leaflets, causing it to degenerate and rendering them incompetent, allowing blood to flow back to the right atrium. It is often explained by infective endocarditis and carcinoid syndrome. Its auscultation, Tricuspid regurgitation is described as a systolic murmur best heard at the LLSB \cite{Thomas:2020}.

\subsubsection{Septal defects}
Septal defects are congenital in nature and are defined as "ruptures" or "discontinuities" in the interatrial septum (atrial septal defects) or in the interventricular septum (ventricular septal defects), allowing the blood to freely flow between them and mix. The auscultation of the atrial septal usually presents as a loud and a wide S1, fixed split S2 heart sound, loudest at the LUSB. Ventricular septal defects often generates a holosystolic murmur, best heard at the apex. Smaller defects are louder and have a harsher quality while large ones are quieter but more symptomatic. Therefore, atrial and ventricular septal defects in children can produce progressively louder murmurs as they close \cite{Thomas:2020}.

\subsubsection{Hypertrophic obstructive cardiomyopathy}
Hypertrophic obstructive cardiomyopathy is an inherited myocardial disease in which the myocardium undergoes hypertrophic changes. During auscultation, hypertrophic obstructive cardiomyopathy presents a systolic ejection murmur, heard best between the apex and the left sternal border (LSB) and becoming louder with Valsalva or abrupt standing maneuvers \cite{Thomas:2020}.

\subsubsection{Patent ductus arteriosus}
In patent ductus arteriosus, a channel fails to close after birth, establishing a shunt that allows oxygenated blood from the aorta to flow back to the lungs via the pulmonary artery. Its auscultation is described as a continuous ``machine-like'' murmur, loudest at the LUSB \cite{Thomas:2020}. 

\section{Available Heart Sound Datasets}
\label{par:datasets}
In this section, we briefly present heart sound datasets that are open to the scientific community. We consider datasets that satisfy the following criteria: a) full availability of data for scrutiny; b) online accessibility; c) relevance to the current study; d) include relevant information regarding the study population (e.g., size, age, and/or gender); and e) include relevant information regarding the audio recordings (e.g., number, duration, or collection spots). With these criteria, we selected six datasets that will form the basis of our analysis. A summary of the selected datasets are presented in Table \ref{DatasetsOpen}. Furthermore, some features from the proposed dataset are also presented in last row of Table \ref{DatasetsOpen}, for comparison reasons. From this table, it is observed that the existing heart sound datasets have significantly different sampling frequencies. While the informative spectra of the PCG is below 1\,kHz (with its power spectrum mainly concentrated below 500\,Hz), the diversity in the sampling frequency of existing datasets is mainly due to the audio devices and computer-based sound cards used for PCG digitization. In fact, since the heart sounds are in the audio frequency range, many research teams have preferred to use high-fidelity audio sound cards, which sample audio signals at standard rates such as 4\,kHz, 8\,kHz, 11025\,Hz, 22050\,Hz, 44100\,Hz, 48\,kHz, etc. In fact, none of these rates are specific to the heart sound, but for historical reasons, they are audio rates supported by standard commercialized audio devices and software \cite[Sec. 4.5]{watkinson2001art}. However, from the signal processing and machine learning perspective, oversampling way above the Nyquist rate (twice the maximum frequency of the desired signal) does not provide additional information regarding the signal and would only increase the audio length and processing load. Based on this fact, with an appropriate anti-aliasing analog filter that avoids ``spectral folding'' effects \cite{oppenheim1999discrete}, a sampling frequency of 2\,kHz to 4\,kHz is fully adequate for all human- and machine-based heart sound diagnosis.
\begin{table*}[t]
  \centering
  \caption{A summary of the currently available open access datasets.\hspace{\textwidth} NPCA : Normal and Pathological Children and Adults; NPC: Normal and Pathological Children; FET : fetus and pregnant; NA : Normal Adults. FH: Fundamental Heart Sounds; NPFM: Normal and Pathological Fetal and Maternal.} 
\begin{tabular}{C{2.5cm}C{0.8cm}C{0.8cm}C{2cm}C{0.8cm}C{1.8cm}C{2cm}C{2cm}C{2cm}C{2cm}C{2cm}C{0.8cm}}
Name                      &   Year  & Type                                        & \# Patients (M/F)       & Age (min;max) years & \# Recordings                                                                          & Duration (min:max) seconds & Sampling rate (Hz) &\# FH Manually Annotated \\
\hline
\hline
Pascal Challenge Dataset \cite{Pascal:2013} &   2011       & NPCA    &     -              & 0-17  &  656       &  1-30                                              & 4000 & 1809          \\
                                                 &                             &        &               &           \\
\hline              
PhysioNet/CinC  \cite{Physionet2}             &   2016       & -                                           & 1297               & -              & 2435             & -             & -         \\
MITHSDB                   & 2007         & NPCA & 121                & -  & 409     & 28-38                              & 44100         &    0       \\
AADHSDB                   & 2015              & NPCA & 151 (93/58)        & -  & 409                                                   & 8                                   & 4000          &   0        \\
AUTHHSDB                  & 2014               & NPCA & 45 (28/17)         & 18-90               &   45                                                                     &         10-122                           & 4000          &  0         \\

TUTHSDB                  & 2013               & NPCA & 44         & -               &   174                                                                     &         15                           & 4000          &  0         \\

UHAHSDB                   & 2013            & NPCA & 55  &    18-40            &  79       &    7-29                                & 8000          &           \\
DLUTHSDB                  & 2012             & NPCA & 509 (280/229)      &    4-88            &  673        &        27.5-312.5                             & 800-22050  &  0         \\
SUAHSDB                   &  2015             & NPCA & 112 (43/69)        &   16-88             & 114                                                                   &  28-38                                  & 8000          &    0       \\
SSHHSDB                   &  -           & NPCA & 35 & -  &         &                        15-69     &         8000          &     0      \\
SUFHSDB                   &  2015            & NPFM    & 225                & 23-35    & 211                                                          & 90                          &         4000-44100    &    0       \\

\hline
HSCT-11  \cite{Spadaccini:2013}                 &   2016       & NPCA & 206 (157/49)       & 6-36      & 412                                                                        & 20-70   & 11025         & 0        \\

\hline
Digiscope    \cite{Oliveira18st}                &   2019       & NPC & 29        & 0-17      & 29                                                                          &  2-20     & 4000         & 612
\\
\hline
Fetal PCG    \cite{Cesarelli:2012}                &   2015        & FET & 26        & 25-35      & 26                                         & -                                    & 333         & 0        \\

\hline

EPHNOGRAM    \cite{EPHNOGRAMDataset}                &   2021         & NA & 24        & 23-29      & 69                                                                         &  30-1800     & 8000         & 0        \\
\hline
\textbf{CirCor DigiScope}
&   \textbf{2021}         & \textbf{NPCA} & \textbf{1568}        & \textbf{0-30}      & \textbf{5282}                                                                         &  \textbf{5-168}     & \textbf{4000}         & \textbf{215780}        \\\hline\hline
\\
\end{tabular}
\label{DatasetsOpen}
\end{table*}
\subsection{Pascal Challenge Database \cite{Pascal:2013} }
This database comprises of two datasets. Dataset A was collected from an unreported size population, through a smartphone application. Dataset B was gathered using a digital stethoscope system deployed in the Maternal and Fetal Cardiology Unit at Real Hospital Português (RHP) in Recife, Brazil. It contains a total of 656 heart sound recordings from an unknown number of patients. The duration of the PCG signals range from 1 to 30 seconds and were collected at a sampling rate of 4000\,Hz.
The sounds were obtained from the apex point of volunteer subjects in the Dataset A. In the Dataset B, sounds were collected from four cardiac auscultation locations on healthy and unhealthy children.
Up to our knowledge, no additional data regarding the auscultation location is provided. The sounds were recorded in clinical and non-clinical environments, and divided into normal, murmur, extra heart sound, and artifact classes in Dataset A. In Dataset B, the sounds were divided into normal, murmur, and extra systole classes. Furthermore, in Dataset B, the positions of fundamental heart sounds were manually annotated by clinicians. 

\subsection{PhysioNet/CinC Challenge 2016 Database  \cite{Physionet2} }
This database merges nine independent databases: the Massachusetts Institute of Technology heart sounds database (MITHSDB), the Aalborg University heart sounds database (AADHSDB), the Aristotle University of Thessaloniki heart sounds database (AUTHHSDB), the Khajeh Nasir Toosi University of Technology heart sounds database (TUTHSDB), the University of Haute Alsace heart sounds database (UHAHSDB), the Dalian University of Technology heart sounds database (DLUTHSDB), the Shiraz University adult heart sounds database (SUAHSDB), the Skejby Sygehus Hospital heart sounds database (SSHHSDB), and the Shiraz University fetal heart sounds database (SUFHSDB).

As part of the PhysioNet/CinC 2016 Challenge, the data has been divided into training and testing sets, and it contains a total of 2435 heart sound records from 1297 patients. The duration of the PCG signals ranges from 8 to 312.5 seconds. Since the data was collected using different devices with different sampling rates, each PCG signal has been downsampled to 2000\,Hz \cite{Clifford:2016} \footnote{The official number of recordings released for the PhysioNet/CinC Challenge was 4430, different from the reported in \cite{Clifford:2016}. This is because the 338 recordings from normal subjects in the DLUTHSDB were divided into several relatively short recordings.}. 
The sounds were obtained from four different auscultation locations (aortic, pulmonary, tricuspid, and mitral) on both healthy (normal) and pathological (abnormal) subjects, with a variety of diseases including heart valve diseases and coronary artery diseases. The training and test sets are unbalanced, with the number of normal records being greater than abnormal records.
The sounds were recorded at clinical and non-clinical environments, and divided into normal, abnormal and unsure classes. 
With the exception of the unsure class, annotations of the positions of fundamental heart sounds were provided for all records. Such annotations were provided by applying an automatic segmentation algorithm \cite{Springer:2016} and manually reviewed and corrected afterwards.


\subsection{HSCT-11 (2016) \cite{Spadaccini:2013}}
Spadaccini \textit{et al.} published an open access database on heart sounds. The dataset was designed to measure the performance of biometric systems, based on auscultation. This dataset is composed of 206 patients and a total of 412 heart sounds, collected from four auscultation locations (mitral, pulmonary, aortic, tricuspid). The data was recorded using the ThinkLabs Rhythm digital electronic stethoscope, with a sampling rate of 11025\,Hz and a resolution of 16\,bits per sample. No information is provided regarding the health condition of each subject.

\subsection{Digiscope \cite{Oliveira18st}}
Oliveira \textit{et al.} released a pediatric dataset composed of 29 heart sounds from 29 patients, ranging in age from six months to 17 years old. The recordings have a minimum, average, and maximum duration of approximately 2, 8, and 20 seconds, respectively. Heart sounds have been collected in Real Hospital Portugu\^{e}s (Recife, Brasil) using a Littmann~3200 stethoscope embedded with the DigiScope Collector \cite{Pereira:2011} technology. The sounds were recorded at 4\,kHz from the Mitral point. Finally, two cardiac physiologists manually annotated the beginning and ending of each fundamental heart sound.

\subsection{Fetal PCG Dataset \cite{Cesarelli:2012}}
Cesarelli \textit{et al.} collected and published 26 fetal heart sounds from different pregnant women, over the last months of their singleton physiological pregnancies. The patients were healthy and aged between 25 to 35 years. The sounds were recorded at 333\,Hz with 8 bits resolution.

\subsection{EPHNOGRAM: A Simultaneous Electrocardiogram and Phonocardiogram Database \cite{EPHNOGRAMDataset,KazemnejadGordanySameni2021}} 
The electro-phono-cardiogram (EPHNOGRAM) project focused on the development of low-cost and low-power devices for recording sample-wise simultaneous electrocardiogram (ECG) and PCG data. The database, has 69 records acquired from 24 healthy young adults aged between 23 and 29 (average: 25.4$\pm$1.9) in 30\,minute stress-test sessions during resting, walking, running and biking conditions (using indoor fitness center equipment). 
The synchronous ECG and PCG channels have been sampled at 8\,kHz with a resolution of 12 bits (with 10.5 effective number of bits). In some records, the environmental audio noises have also been recorded through additional auxiliary audio channels, which can be used to enhance the main PCG channel signal quality through signal processing. This data is useful for simultaneous multi-modal analysis of the ECG and PCG, as it provides interesting insights about the inter-relationship between the mechanical and electrical mechanisms of the heart, under rest and physical activity. No manual segmentation has been made, but a MATLAB code is provided that automatically and accurately detect all the R-peaks from the ECG. Such peaks can be used to locate the S1 and S2 components of the PCG.


\section{Methods}
\label{par:proposed}
\begin{figure}
    \centering
    \includegraphics[width=\columnwidth]{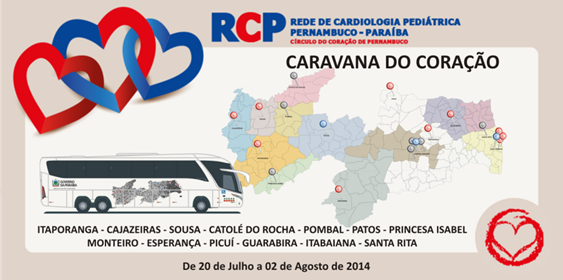}
    \caption{The CC2014 screening campaign road map. The participants were from: Itaporanga, Cajazeiras, Sousa, Catolé da Rocha, Pombal, Patos, Princesa Isabel, Monteiro, Esperança, Picuí, Guarabina, Itabaina and Santa Rita }
    \label{Caravana_do_Coracao_2014_map}
\end{figure}

\begin{figure}
    \centering
    \includegraphics[width=\columnwidth]{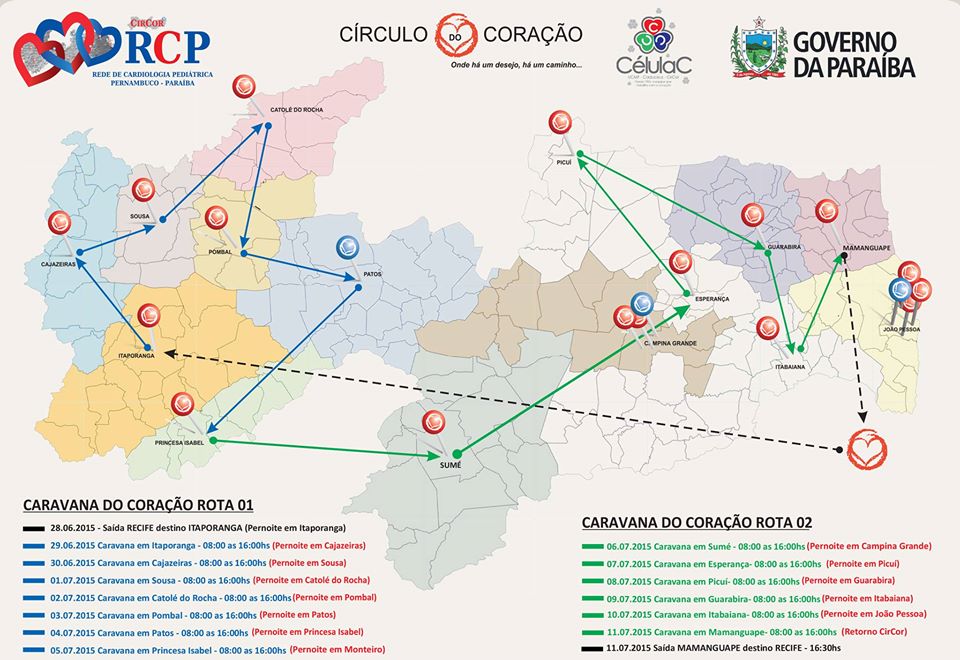}
    \caption{The CC2015 screening campaign road map. The participants were from: Itaporanga, Cajazeiras, Sousa, Catolé da Rocha, Pombal, Patos, Princesa Isabel, Sumé, Esperança, Picuí, Guarabina, Itabaina and Mamangrape }
    \label{Caravana_do_Coracao_2015_map}
\end{figure}

The presented dataset was collected as part of two mass screening campaigns, referred to as ``Caravana do Coração'' (Caravan of the Heart) campaigns, conducted in the state of Paraíba, Brazil between July and August 2014 (CC2014) and June and July 2015 (CC2015), see Figures \ref{Caravana_do_Coracao_2014_map} and \ref{Caravana_do_Coracao_2015_map}. The data collection was approved by the 5192--Complexo Hospitalar HUOC/PROCAPE institution review board, under Real Hospital Portugues de Beneficiencia em Pernambuco request. The CC2014 and CC2015 screening campaigns were promoted by a non-governmental organization (NGO) from Pernambuco, ``Círculo do Coração'' (Heart Circle - CirCor) and funded by the Health Secretary of the neighbouring State, Paraíba. These caravans occurred during a seven-year partnership program, which established the Pediatric Cardiology Network, an integral line of care --- from screening to surgery and follow-up ---  for under-served children with heart diseases in Paraíba.
The collected dataset is provided online on PhysioNet \cite{Oliveira:2022}. Note that, in order to manage the organization of public data challenges on murmur grading and classification, only 70\% of the gathered dataset was publicly released in the PhysioNet repository (randomly selected through stratified random sampling). The remaining 30\%, which is currently held as a private repository for test purposes (as unseen data), it will be released on the same PhysioNet repository after the corresponding data challenge. The procedure of data gathering and the properties of the dataset are detailed in the following sections.

\subsection{The Caravana do Coração Campaigns}

The studied population included participants who volunteered for screening within the study period. Patients younger than 21 years of age with a parental signed consent form (where appropriate) were included. A total of 2,061 participants attended the 2014 and 2015 campaigns, with 493 participants being excluded for not meeting the eligibility criteria. Furthermore, 116 patients attended both screening campaign. In our database, these patients are identified and highlighted. 

All participants completed a socio-demographic questionnaire and subsequently underwent clinical examination (anamnesis and physical examination), nursing assessment (physiological measurements), and cardiac investigation (chest radiography, electrocardiogram, and echocardiogram). Data quality assessment was performed and all entries were screened for incorrectly entered or measured values, inconsistent data, or for the presence of outliers, and omitted in case of any inconsistency. The resulting entries were then compiled in a spreadsheet to reflect the socio-demographic and clinical variables used in our dataset. Subsequently, an electronic auscultation was performed and audio samples from four typical auscultation spots were collected; all samples were collected by the same operator for the duration of the screening, in a real clinical setting. Two independent cardiac physiologists individually assessed the resulting PCG audio files for signal quality. As a result, 119 participants did not meet the required signal quality standards. In another words, the patient recordings do not allow a safe and trustworthy murmur characterization and description, see Table \ref{CC2014_2015Indicators} .



\subsection{Demographic and Clinical Information}
The collected dataset includes 1568 participants, 787 (50.2\%) were male and 781 (49.8\%) were female. Regarding age categories, following the National Institute of Child Health and Human Development (NICHD) pediatric terminology \cite{Williams:2012}, 988 are children (63.0\%), 311 infants (19.80\%),  127 adolescents (8.1\%), 9 young adults (0.6\%), 11 neonates (0.7\%) and 110 pregnant women (8.1\%); in 12 patients no age data is provided (0.8\%). Finally, with regards to ethnicity, 1298 participants are mixed race (82.8\%), 249 (15.9\%) white, and 1.4\% of other ethnic backgrounds. Table \ref{GenderAgeDistribution} summarizes the population demography. 

\begin{table}[tb]
\centering
\caption{Gender,age group, child's race and mother's race distribution from the CC2014 and CC2015 screening campaigns.}
\begin{tabular}{rccr}
\hline
\multicolumn{1}{l}{}                       & \multicolumn{1}{l}{CC2014 (\%)} & \multicolumn{1}{l}{CC2015 (\%)} & \textbf{TOTAL (\%)}   \\ \hline
\multicolumn{1}{l}{\textit{Gender}}        & \multicolumn{1}{l}{}        & \multicolumn{1}{l}{}        & \textbf{}            \\
Male                                       & 325 (49.8)                  & 462 (50.5)                  & \textbf{787 (50.2)}  \\
Female                                     & 328 (50.2)                  & 453 (49.5)                  & \textbf{781 (49.8)}  \\
\multicolumn{1}{l}{}                       & \multicolumn{1}{l}{}        & \multicolumn{1}{l}{}        & \multicolumn{1}{l}{} \\
\multicolumn{1}{l}{\textit{Age Group}}     & \multicolumn{1}{l}{}        & \multicolumn{1}{l}{}        & \multicolumn{1}{l}{} \\
Child                                      & 405 (62.0)                  & 583 (63.7)                  & \textbf{988 (63.0)}  \\
Infant                                     & 126 (19.3)                  & 185 (20.2)                  & \textbf{311 (19.8)}  \\
Pregnant                                      & 57 (8.7)                    & 53 (5.8)                    & \textbf{110 (7.0)}   \\
Adolescent                                 & 51 (7.8)                    & 76 (8.3)                    & \textbf{127 (8.1)}   \\
Young adult                                & 5 (0.8)                     & 4 (0.4)                     & \textbf{9 (0.6)}     \\
Neonate                                    & 3 (0.5)                     & 8 (0.9)                     & \textbf{11 (0.7)}    \\
No info                                    & 6 (0.9)                     & 6 (0.7)                     & \textbf{12 (0.8)}    \\
\multicolumn{1}{l}{}                       & \multicolumn{1}{l}{}        & \multicolumn{1}{l}{}        & \multicolumn{1}{l}{} \\
\multicolumn{1}{l}{\textit{Race (child)}}  & \multicolumn{1}{l}{}        & \multicolumn{1}{l}{}        & \multicolumn{1}{l}{} \\
Mixed Race                                 & 492 (75.3)                  & 806 (88.1)                  & \textbf{1298 (82.8)} \\
White                                      & 151 (23.1)                  & 98 (10.7)                   & \textbf{249 (15.9)}  \\
Black                                      & 9 (1.4)                     & 11 (1.2)                    & \textbf{20 (1.3)}    \\
Asian                                      & 1 (0.2)                     & 0 (0.0)                     & \textbf{1 (0.1)}     \\
\multicolumn{1}{l}{}                       & \multicolumn{1}{l}{}        & \multicolumn{1}{l}{}        & \textbf{}            \\
\multicolumn{1}{l}{\textit{Race (mother)}} & \multicolumn{1}{l}{}        & \multicolumn{1}{l}{}        & \textbf{}            \\
Mixed Race                                 & 389 (59.6)                  & 705 (77.0)                  & \textbf{1094 (69.8)} \\
White                                      & 240 (36.8)                  & 195 (21.3)                  & \textbf{435 (27.7)}  \\
Black                                      & 24 (3.7)                    & 14 (1.5)                    & \textbf{38 (2.4)}    \\
Asian                                      & 0 (0.0)                     & 1 (0.1)                     & \textbf{1 (0.1)}     \\ \hline
\end{tabular}
\label{GenderAgeDistribution}
\end{table}

The mean age ($\pm$ standard deviation) of the participants is $73.4\pm 0.1$ months, ranging from 0.1 to 356.1 months, as summarized in Table \ref{AgeStatistics}. 

\begin{table}[tb]
\centering
\caption{Age statistics of the participants in months}
\begin{tabular}{rccc}
\hline
\multicolumn{1}{l}{} (months) & CC2014 & CC2015 & \textbf{TOTAL} \\\hline
Mean                            & 74.7   & 72.5   & \textbf{73.4}  \\
Median                        & 78.4   & 70     & \textbf{72.1}  \\
Standard deviation                         & 50.4   & 50.3   & \textbf{50.3}  \\
Minimum                             & 0.1    & 0.1    & \textbf{0.1}   \\
Maximum                             & 217.8  & 356.1  & \textbf{356.1} \\
\hline
\end{tabular}
\label{AgeStatistics}
\end{table}

The majority of the participants attended the study with no formal indication (444; 27.0\%), while (305; 18.5\%) presented for follow-up of a previously diagnosed cardiopathy. Moreover, 223 participants (13.5\%) attended the screening campaign to investigate the evolution of previously identified murmurs. A total of 647 single or multiple diagnosis were confirmed, the most frequent being simple congenital cardiopathy (30.2\%) and acquired cardiopathy (3.3\%), with 65 (3.9\%) diagnosis of complex congenital cardiopathy also being established. A total of 834 (53.2\%) participants were referred for follow-up, 27 (1.2\%) were referred for additional testing, and 35 (2.2\%) had indication for surgery/intervention. 575 participants (36.7\%) were discharged after screening. Table~\ref{IndicationDiagnosisPlanning2014and2015} summarizes these statistics. 

\begin{table}[tb]
\centering
\caption{Indications, diagnosis and plan for patients presenting to the 2014 and 2015 campaigns.}
\resizebox{\columnwidth}{!}{
\begin{tabular}{rccr}
\hline
\multicolumn{1}{l}{}                    & \multicolumn{1}{l}{ CC2014 (\%)}            & \multicolumn{1}{l}{ CC2015 (\%)}            & \textbf{TOTAL (\%)}    \\ \hline
\multicolumn{1}{l}{\textit{Indication}} & \multicolumn{1}{l}{}                     & \multicolumn{1}{l}{}                     & \textbf{}             \\
Screening: no formal indication         & 188 (27.1)                              & 256 (26.9)                               & \textbf{444 (27.0)}   \\
Murmur                                  & 133 (19.2)                               & 90 (9.5)                                 & \textbf{223 (13.6)}   \\
Previously diagnosed cardiopathy        & 79 (11.4)                                & 226 (23.7)                               & \textbf{305 (18.5)}   \\
Screening: fetal cardiac assessment     & 57 (8.2)                                 & 55 (5.8)                                 & \textbf{112 (6.8)}    \\
Rate/rhythm/conduction disturbance      & 38 (5.5)                                 & 51 (5.4)                                 & \textbf{89 (5.4)}     \\
Chest Pain                              & 33 (4.8)                                 & 29 (3.0)                                 & \textbf{62 (3.8)}     \\
Family history of cardiopathy           & 30 (4.3)                                 & 28 (2.9)                                 & \textbf{58 (3.5)}     \\
Cyanosis                                & 18 (2.6)                                 & 15 (1.5)                                 & \textbf{33 (2.0)}     \\
Fatigue                                 & 17 (2.5)                                 & 14 (1.5)                                 & \textbf{31 (1.9)}     \\
Syncope                                 & 19 (2.7)                                 & 11 (1.2)                                 & \textbf{30 (1.8)}     \\
Post-op assessment cardiac surgery      & 14 (2.0)                                 & 79 (8.3)                                 & \textbf{93 (5.7)}     \\
Hypertension                            & 10 (1.4)                                 & 8 (0.8)                                  & \textbf{18 (1.1)}     \\
Genetic syndrome                        & 5 (0.7)                                  & 15 (1.6)                                 & \textbf{20 (1.2)}     \\
Suspected rheumatic fever               & 24 (3.5)                                 & 10 (1.1)                                 & \textbf{34 (2.0)}     \\
Seizure                                 & 8 (1.2)                                  & 7 (0.7)                                  & \textbf{15 (0.9)}     \\
Dyspnea                                 & 4 (0.6)                                  & 15 (1.6)                                 & \textbf{19 (1.2)}     \\
Inflammatory syndrome                   & 2 (0.3)                                  & 2 (0.2)                                  & \textbf{4 (0.2)}      \\
Pre-op assessment non-cardiac surgery   & 1 (0.1)                                  & 3 (0.3)                                  & \textbf{4 (0.2)}      \\
Previous cardiorespiratory arrest       & 1 (0.1)                                  & 0 (0.0)                                  & \textbf{1 (0.1)}     \\
Myocarditis                             & 1 (0.1)                                  & 0 (0.0)                                  & \textbf{1 (0.1)}     \\
Screening: other not specified          & 11 (1.6)                                 & 5 (0.5)                                  & \textbf{16 (1.0)}     \\
Recurrent tonsillitis                   & 0 (0.0)                                  & 2 (0.2)                                  & \textbf{2 (0.1)}      \\
Overweight                              & 0 (0.0)                                  & 3 (0.3)                                  & \textbf{3 (0.2)}      \\
Chest x-ray changes                     & 0 (0.0)                                  & 3 (0.3)                                  & \textbf{3 (0.2)}      \\
Rheumatic fever                         & 0 (0.0)                                  & 25 (2.6)                                 & \textbf{25 (1.5)}     \\
\textbf{TOTAL}                          & \multicolumn{1}{r}{\textbf{693 (100.0)}} & \multicolumn{1}{r}{\textbf{952 (100.0)}} & \textbf{1645 (100.0)} \\
\multicolumn{1}{l}{}                    & \multicolumn{1}{l}{}                     & \multicolumn{1}{l}{}                     & \textbf{}             \\
\multicolumn{1}{l}{\textit{Diagnosis}}  & \multicolumn{1}{l}{}                     & \multicolumn{1}{l}{}                     & \textbf{}             \\ \hline
Simple Congenital Cardiopathy           & 195 (26.7)                               & 310 (32.8)                               & \textbf{505 (30.2)}   \\
Acquired Cardiopathy                    & 36 (5.0)                                 & 19 (2.0)                                 & \textbf{55 (3.3)}     \\
Complex Congenital Cardiopathy          & 21 (2.9)                                 & 44 (4.7)                                 & \textbf{65 (3.9)}     \\
Arrhythmia                              & 10 (1.4)                                 & 12 (1.3)                                 & \textbf{22 (1.3)}     \\
Normal                                  & 176 (24.2)                               & 497 (52.5)                               & \textbf{673 (40.2)}   \\
No info                                 & 290 (39.9)                               & 64 (6.8)                                 & \textbf{354 (21.1)}   \\
\textbf{TOTAL}                          & \multicolumn{1}{r}{\textbf{728 (100.0)}} & \multicolumn{1}{r}{\textbf{946 (100.0)}} & \textbf{1674 (100.0)} \\
\multicolumn{1}{l}{}                    & \multicolumn{1}{l}{}                     & \multicolumn{1}{l}{}                     & \multicolumn{1}{l}{}  \\
\multicolumn{1}{l}{\textit{Plan}}       & \multicolumn{1}{l}{}                     & \multicolumn{1}{l}{}                     & \multicolumn{1}{l}{}  \\ \hline
Follow-up                               & 270 (41.3)                               & 564 (61.6)                               & \textbf{834 (53.2)}   \\
Discharged                              & 278 (42.6)                               & 297 (32.5)                               & \textbf{575 (36.7)}   \\
Additional testing required             & 26 (4.0)                                 & 1 (0.1)                                  & \textbf{27 (1.2)}     \\
Indication for surgery or intervention  & 28 (4.3)                                 & 7 (0.8)                                  & \textbf{35 (2.2)}     \\
No information                          & 51 (7.8)                                 & 46 (5.0)                                 & \textbf{97 (6.2)}     \\
\textbf{TOTAL}                          & \multicolumn{1}{r}{\textbf{653 (100.0)}} & \multicolumn{1}{r}{\textbf{915 (100.0)}} & \textbf{1568 (100.0)} \\ \hline
\end{tabular}
}
\label{IndicationDiagnosisPlanning2014and2015}
\end{table}

Regarding clinical presentation, 1401 participants (89.4\%) presented a good general condition, with the majority of patients being eupneic (90.2\%) and having normal perfusion (90.7\%).

\begin{figure}[tb]
\centering
\begin{tikzpicture}
\begin{axis}[
ymax=200,
ymin=40,
ylabel={Beats per Minute},
xmax=1,
xmin = 0,
boxplot/draw direction=y,
cycle list={{ultra thick}, {ultra thin}},
legend cell align={right}, 
        legend pos= south east,
        legend image post style={sharp plot},
boxplot={
       draw position={
                1/5 + floor(\plotnumofactualtype/2)
                  + 1/2*fpumod(\plotnumofactualtype,2)
            },
            box extend=0.1,},
xtick={0,0.40,1.0},
x tick label as interval,
xticklabels={CC2014, CC2015},
		label style={font=\small},
        x tick label style={font=\small},
]      
]
\addplot+[mark = x, mark options = {mark color=grey},boxplot]
table[y index=0] {HeartRateCC2014.txt};

\addplot+[mark = x, mark options = {mark color=grey},boxplot]
table[y index=0] {HeartRateCC2015.txt};

\end{axis}
\end{tikzpicture}
\caption{The heart rate distribution of the CC2014 (on the left) and CC2015 (on the right) screening campaign.} \label{HeartRateDistribution}
\end{figure}
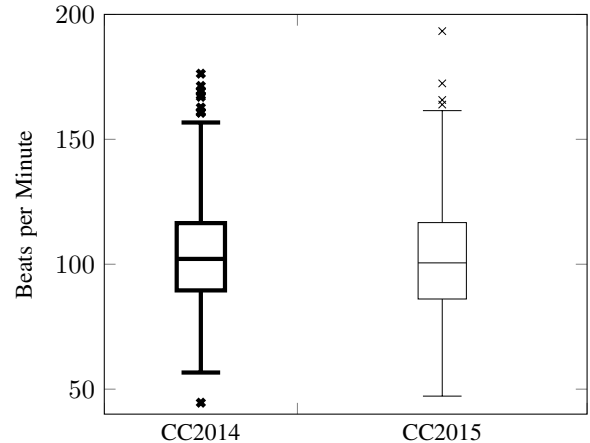

The mean weight and height ($\pm$ standard deviation) of the sample is $24\pm15$ kg and $111\pm29$ cm respectively, with a mean BMI of $18\pm7$. The mean heart rate is $102\pm20$ bpm, ranging between 47 bpm  and 193 bpm. A Mann-Whitney U hypothesis test on the heart rate distributions was performed, a p-value of 0.04 was observed, see Figure \ref{HeartRateDistribution} for more details. The mean oxygen saturation (measured in the arm) is $95\% \pm5\%$. With regards to tympanic temperature, the average temperature is $36.7\pm0.4$ C, ranging from $34.7$ C to $39.2$ C. Finally, the average systolic and diastolic blood pressure is $91 \pm 17$ mmHg and $55\pm 11$ mmHg, respectively. The maximum systolic and diastolic blood pressure is, respectively, 170 mmHg and 101 mmHg.

\subsection{Heart Sounds and Annotations}
The collected dataset includes a total number of 215780 heart sounds, 103853 heart sounds (51 945 S1 and 51 908 S2 waves) from CC2014 and 111927 (56449 S1 and 55478 S2 waves) from the CC2015. 

In the CC2014 screening campaign, 540 recordings were collected from the Aortic point, 497 from the Pulmonary point, 603 from the Mitral point, 461 from the Tricuspid point and 5 from an unreported point. Between 1 to 10 records have been recorded per patient, with an average of 3.2 recordings per patient. In the CC2015 screening campaign, 817 recordings were collected from the Aortic point, 793 from the Pulmonary point, 812 from the Mitral point, 754 from the Tricuspid point, and one extra sound from an unreported point. Overall, between 1 to 4 records exist per patient, with an average of 3.5 recordings per patient. \\

\begin{figure}[tb]
    \centering
    \includegraphics[width=\columnwidth]{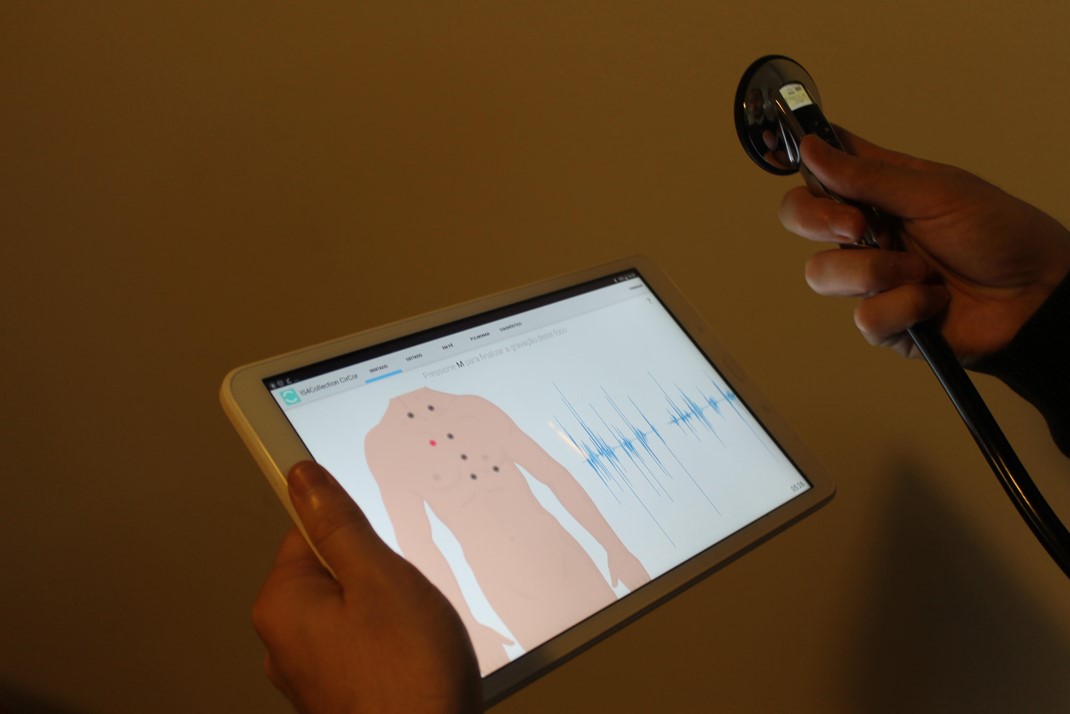}
    \caption{The DigiScope Collector technology. The signals are collected using a Littmann 3200 stethoscope. Afterwards, the signals are transmitted and visualized (near real time) in a tablet device.   }
    \label{DigiScopeCollector}
\end{figure}

\begin{figure*}
\centering

  \includegraphics[width=0.3\linewidth]{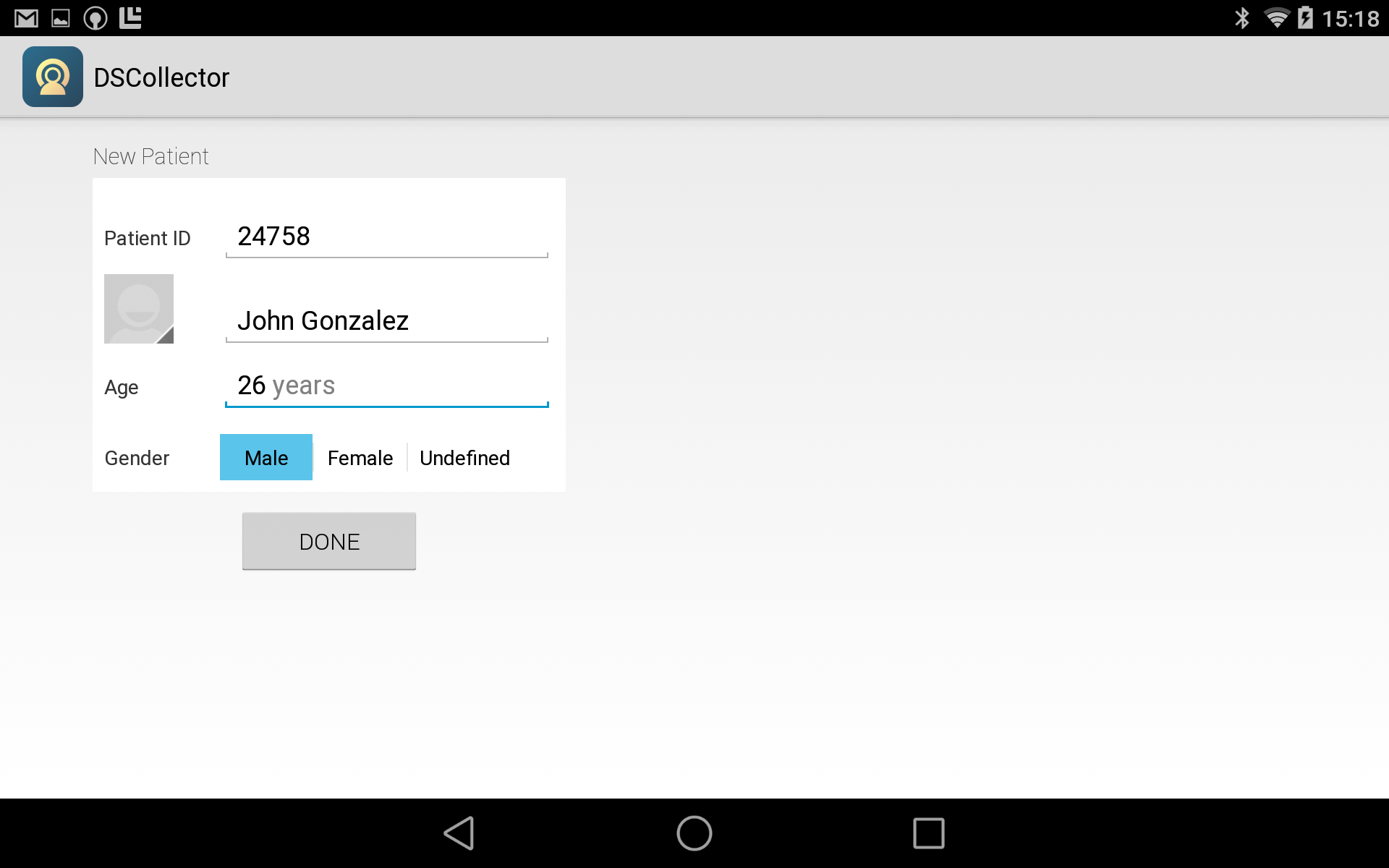} \hfill
  \includegraphics[width=0.3\linewidth]{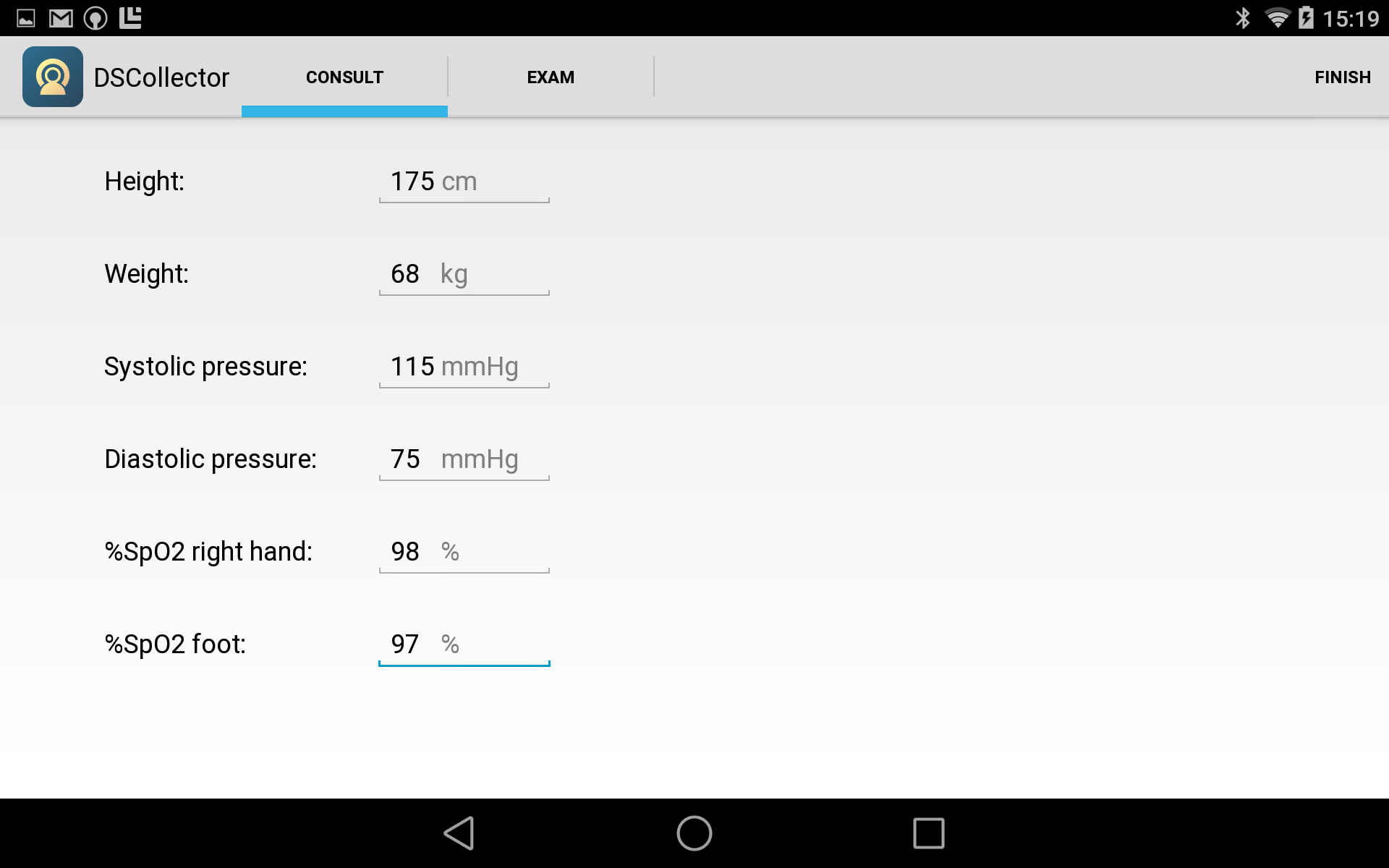}\hfill
  \includegraphics[width=0.3\linewidth]{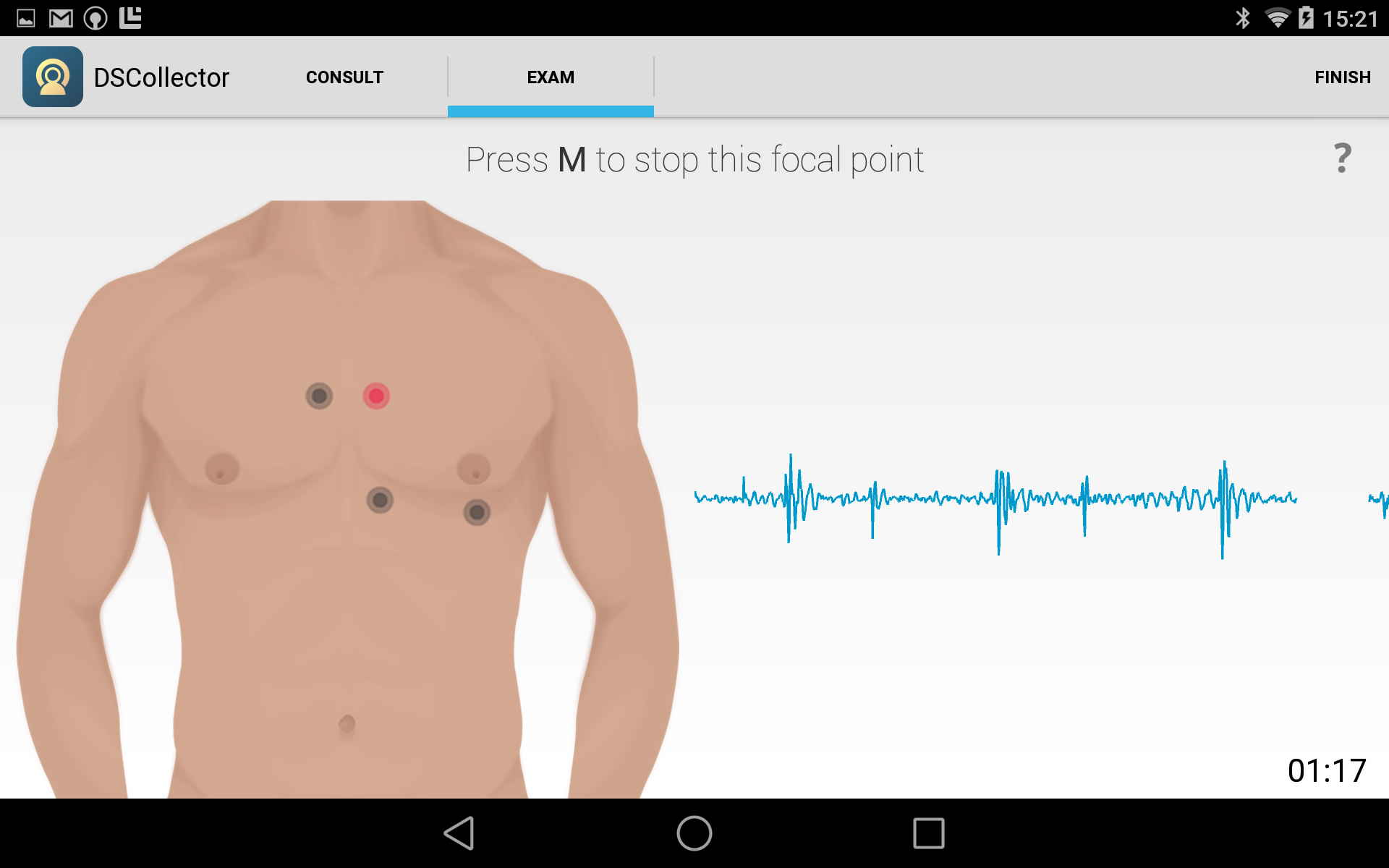}\hfill
\caption{Screenshots of the DigiScope Collector's Graphical User Interface (GUI), during the different data acquisition stages. In the left panel, the patient creation layout; in the middle panel, clinical data from the patient is inserted; in the right panel, acquisition of a heart sound signal, from the Pulmonary spot. For more details see \cite{Gomes:2016}.}
\label{DigiScopeCollectorArq}
\end{figure*}

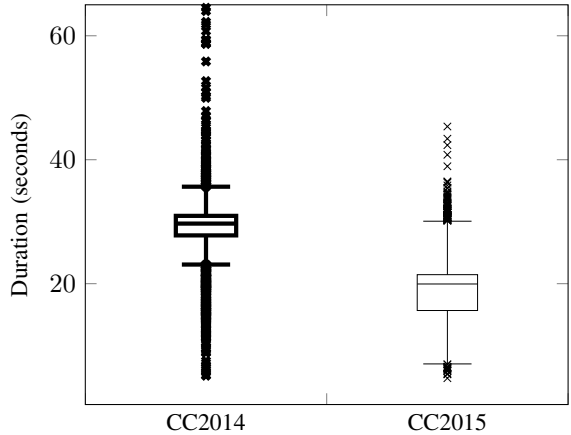
\begin{figure}[tb]
\centering
\begin{tikzpicture}
\begin{axis}[
ymax=65,
ymin=0.5,
ylabel={Duration (seconds)},
xmax=8,
xmin =0,
boxplot/draw direction=y,
cycle list={{ultra thick}, {ultra thin}},
legend cell align={right}, 
        legend pos= south east,
        legend image post style={sharp plot},
boxplot={
       draw position={
               2 + floor(\plotnumofactualtype)
                  + 3*fpumod(\plotnumofactualtype,2)
            },
            box extend=1,},
xtick={0,4, 8},
x tick label as interval,
xticklabels={ CC2014, CC2015},
		label style={font=\small},
        x tick label style={font=\small},
]      
]

\addplot+[mark = x, mark options = {mark color=grey},boxplot]
table[y index=0] {Signal_Length_Distribution_Raw_Folder.txt};

\addplot+[mark = x, mark options = {mark color=grey},boxplot]
table[y index=0] {Signal_Length_Distribution_Raw_Folder_CC3.txt};

\end{axis}
\end{tikzpicture}
\caption{The CC2014 and CC2015 screening campaign Folder distribution. Folder CC2014: min=5.3 seconds; max=80.4 seconds; average = 28.7 seconds; total = 144.4 hours of recording.  Folder CC2015: min=4.8 seconds; max=45.4 seconds; average = 19.0 seconds; total = 167.7 hours of recording.
} \label{DurationDistribution}
\end{figure}

The heart sound signals were collected using a Littmann 3200 stethoscope embedded with the DigiScope Collector \cite{Pereira:2011} technology, an illustration is presented in Figure \ref{DigiScopeCollector}. A diagram of the graphical user interface is presented in Figure \ref{DigiScopeCollectorArq}. The signal was sampled at 4KHz and with a 16-bits resolution. Furthermore. the heart sound signals are normalized within the $[-1,1]$ range. The PCG files from the CC2014 and CC2015 campaigns, had an average duration of 28.7 seconds and 19.0 seconds, respectively. A Mann-Whitney U hypothesis test on the signal duration distributions was performed, a p-value $\ll$ 0.001 was observed. For more details refer to Figure~\ref{DurationDistribution}.

Murmurs were present in 305 patients within the collected dataset. Out of these, 294 patients had only a systolic murmur, 1 patient had only a diastolic murmur, and 9 patients had both systolic and diastolic murmurs, as summarized in Table~\ref{CC2014_2015Indicators}.

\begin{table}[tb]
\centering
\caption{Distribution of the most important indicators of the CC2014 and CC2015 screening campaign.
}
\resizebox{0.5\textwidth}{!}{
\begin{tabular}{lrccl}
\hline
                              & \multicolumn{1}{l}{}       & \multicolumn{1}{l}{CC2014} & \multicolumn{1}{l}{CC2015} & TOTAL         \\\hline
\textit{Murmur}               & \multicolumn{1}{l}{}       & \multicolumn{2}{c}{n(\%)}                            &               \\
                              & Absent                     & 506 (77.5)               & 638 (69.7)               & \textbf{1144 (73.0)} \\
                              & Present                    & 109 (16.7)               & 196 (21.3)               & \textbf{305 (19.5)}  \\
                              & Low Quality                & 38 (5.8)                 & 81 (8.9)                 & \textbf{119 (7.6)}  \\
\textbf{TOTAL}                & \multicolumn{1}{l}{}       & \multicolumn{1}{l}{}     & \multicolumn{1}{l}{}     & \textbf{1568 (100.0)} \\
\textit{Murmur timing}        & \multicolumn{1}{l}{}       & \multicolumn{1}{l}{}     & \multicolumn{1}{l}{}     &               \\
                              & Early-systolic             & 35 (30.7)                & 68 (34.3)                & \textbf{103 (32.8)}  \\
                              & Mid-systolic               & 6 (5.3)                  & 21 (10.5)                & \textbf{27 (8.6)}   \\
                              & Holosystolic               & 68 (59.6)                & 103 (51.5)               & \textbf{171 (54.5)}  \\
                              & Late-systolic              & 0 (0.0)                  & 3 (1.5)                  & \textbf{3 (1.0)}    \\
                              & Early-diastolic            & 5 (4.4)                  & 3 (1.5)                  & \textbf{8 (2.5)}    \\
                              & Holodiastolic              & 0 (0.0)                  & 1 (0.5)                  & \textbf{1 (0.3)}    \\
                              & Mid-diastolic              & 0 (0.0)                  & 1 (0.5)                  & \textbf{1 (0.3)}    \\
\textbf{TOTAL}                & \multicolumn{1}{l}{}       & \multicolumn{1}{l}{}     & \multicolumn{1}{l}{}     & \textbf{314 (100.0)}  \\
\textit{Murmur pitch}         & \multicolumn{1}{l}{}       & \multicolumn{1}{l}{}     & \multicolumn{1}{l}{}     &               \\
                              & High                       & 21 (18.4)                & 54 (27.0)                & \textbf{75 (23.9)}   \\
                              & Medium                     & 41 (36.0)                & 53 (26.5)                & \textbf{94 (29.9)}   \\
                              & Low                        & 52 (45.6)                & 93 (46.5)                & \textbf{145 (46.2)}  \\
\textbf{TOTAL}                & \multicolumn{1}{l}{}       & \multicolumn{1}{l}{}     & \multicolumn{1}{l}{}     & \textbf{314 (100.0)}  \\
\textit{Murmur grading}       & \multicolumn{1}{l}{}       & \multicolumn{1}{l}{}     & \multicolumn{1}{l}{}     &               \\
                             \textit{systolic} & I/VI                       & 77 (67.5)                & 99 (49.5)                & \textbf{176 (56.1)}  \\
                              & II/VI                      & 10 (8.8)                 & 36 (18.0)                & \textbf{46 (14.6)}   \\
                              & III/VI                     & 22 (19.3)                & 60 (30.0)                & \textbf{82 (26.1)}   \\
                             \textit{Murmur grading } & I/IV                       & 2 (1.8)                  & 5 (2.5)                  & \textbf{7 (2.0)}    \\
                             \textit{diastolic} & II/IV                      & 2 (1.8)                  & 0 (0.0)                  & \textbf{2 (0.6)}    \\
                              & III/IV                     & 1 (0.9)                  & 0 (0.0)                  & \textbf{1 (0.3)}    \\
\textbf{TOTAL}                & \multicolumn{1}{l}{}       & \multicolumn{1}{l}{}     & \multicolumn{1}{l}{}     & \textbf{314 (100.0)}  \\
\textit{Murmur shape}         & \multicolumn{1}{l}{}       & \multicolumn{1}{l}{}     & \multicolumn{1}{l}{}     &               \\
                              & Crescendo                  & 1 (0.9)                  & 4 (2.0)                  & \textbf{5 (1.6)}    \\
                              & Decrescendo                & 29 (25.4)                & 34 (17.0)                & \textbf{63 (20.1)}   \\
                              & Diamond                    & 14 (12.3)                & 44 (22.0)                & \textbf{58 (18.5)}   \\
                              & Plateau                    & 70 (61.4)                & 118 (59.1)               & \textbf{188 (59.9)}  \\
\textbf{TOTAL}                & \multicolumn{1}{l}{}       & \multicolumn{1}{l}{}     & \multicolumn{1}{l}{}     & \textbf{314 (100.0)}  \\
\textit{Murmur quality}       & \multicolumn{1}{l}{}       & \multicolumn{1}{l}{}     & \multicolumn{1}{l}{}     &               \\
                              & Blowing                    & 48 (42.1)                & 97 (48.5)                & \textbf{145 (46.2)}  \\
                              & Harsh                      & 62 (54.4)                & 103 (51.5)               & \textbf{165 (52.5)}  \\
                              & Musical                    & 4 (3.5)                  & 0 (0.0)                  & \textbf{4 (1.3)}    \\
\textbf{TOTAL}                & \multicolumn{1}{l}{}       & \multicolumn{1}{l}{}     & \multicolumn{1}{l}{}     & \textbf{314 (100.0)}  \\
\multicolumn{2}{l}{\textit{Highest murmur intensity location}} & \multicolumn{1}{l}{}     & \multicolumn{1}{l}{}     &               \\
                              & {Aortic}            & 10 (9.2)                 & 30 (15.3)                & \textbf{40 (13.1)}   \\
                              & {Mitral}            & 32 (29.4)                & 31 (15.8)                & \textbf{ 63 (20.7)}   \\
                              & {Pulmonary}          & 43 (39.4)                & 68 (34.7)                & \textbf{111 (36.4)}  \\
                              & {Tricuspid}         & 24 (22.0)                & 67 (34.2)                & \textbf{91 (29.8)}   \\
\textbf{TOTAL}                & \multicolumn{1}{l}{}       & \multicolumn{1}{l}{}     & \multicolumn{1}{l}{}     & \textbf{305 (100.0)} \\

\multicolumn{2}{l}{\textit{Murmur per location}} & \multicolumn{1}{l}{}     & \multicolumn{1}{l}{}     &               \\
                              & {Aortic}            & 66 (23.3)                 & 128 (22.7)                & \textbf{194 (22.9)}   \\
                              & {Mitral}            & 80 (28.3)                & 147 (26.1)                & \textbf{227 (26.8)}   \\
                              & {Pulmonary}          & 71 (25.1)                & 142 (25.2)                & \textbf{213 (25.2)}  \\
                              & {Tricuspid}         & 64 (22.6)                & 145 (25.8)                & \textbf{209 (24.7)}   \\
                             & {Other}         & 2 (0.7)                & 1 (0.2)                &   \textbf{3 (0.4)}   \\
\textbf{TOTAL}                & \multicolumn{1}{l}{}       & \multicolumn{1}{l}{}     & \multicolumn{1}{l}{}     & \textbf{846 (100.0)} 
\\\hline
\end{tabular}
}
 \label{CC2014_2015Indicators}
\end{table}

\section{Data Labelling}
\label{par:methods}
The acquired audio samples were automatically segmented using the three algorithms proposed by \cite{Springer:2016}, \cite{Oliveira18st} and \cite{Renna19}. These algorithms detect and identify the fundamental heart sounds (S1 and S2 sounds) and their corresponding boundaries. As a result, a set  of  annotation  recommendations for each heart sound signal were provided to the two  cardiac physiologists. The two cardiac physiologists independently inspected the resulting algorithms’ output on mutually  exclusive data. In other words, they did not  over-read  each other's annotations. Nevertheless, experts were free to use or neglect the generated recommendations. The expert first inspected the automatic segmentation annotations and re-annotated the misdetections. For each heart sound signal, the  expert  started  by  randomly  selecting one of the corresponding annotation recommendation, and one of two actions follows:
\begin{itemize}
    \item In  case  of  agreement  with  the  algorithm’s  segmentation output, the audio file and its corresponding annotation file were included in the dataset.
    \item In  case  of  disagreement,  the  expert  randomly  opens another recommended annotation and the procedure is repeated.
\end{itemize}
If the expert disagreed with all recommended annotations, then he/she proceeded to manually segment at least five heartbeat cycles. The new audio file and its corresponding annotation file were then saved.

Labels were retains for sections of the data for which the cardiac physiologists indicated were high quality representative segments. The remainder of the signal may include both low and high quality data. In this way, the user is free to use (or not) the suggested time window, where the quality of signal was manually inspected and the automated labels were  validated.

This methodology was applied to both the 2014 and 2015 screening campaigns, resulting in two independent folders (CC2014, CC2015), as present on the project repository on PhysioNet \cite{Oliveira:2022}.

The audio and the corresponding segmentation annotation file names are in \texttt{ABCDE\_XY.wav} format, where \texttt{ABCDE} is a numeric patient identifier and \texttt{XY} is one of the following codes, corresponding to the auscultation location where the PCG was collected on the body surface: \texttt{PV} corresponds to the Pulmonary point; \texttt{TV} corresponds to the Tricuspid point; \texttt{AV} corresponds to the Aortic point, \texttt{MV} corresponds to the Mitral point, and finally \texttt{PhC} for any other auscultation location. If more than one recording exists per auscultation location, an integer index proceeds the auscultation location code in the file name, i.e, \texttt{ABCDE\_nXY.wav}, where \texttt{n} is an integer. Furthermore, each audio file has its own corresponding annotation segmentation file. The segmentation annotation file is composed by three distinct columns: the first column corresponds to the time instance, where the wave was detected for the first time; the second column correspond to the time instance, where the wave was detected for the last time; the third column correspond to an identifier, that uniquely identifies the detected wave.

All the collected heart sound records were also screened for presence of murmurs at each auscultation location. Each murmur was classified (at the auscultation point where it is most audible according to the details in Section~\ref{sec:clinical}) according to its timing (early-, mid-,  and late- systolic/diastolic) \cite{Owen2015}, shape (crescendo, decrescendo, diamond, plateau), pitch (high,  medium, low), quality (blowing, harsh, musical) \cite{Owen2015}, and grade (according  to  Levine’s  scale \cite{Levine:1933}). Since not all patients have auscultation sounds recorded from all the four main auscultation locations, 
the strategy adopted to provide grading annotations is described in Table \ref{GradingTable}

\begin{table}[tb]
\caption{The adopted grading scale}
\begin{tabular}{p{0.1\textwidth}p{0.35\textwidth}}
\textbf{GRADE} & \textbf{Description} \\
 Grade I & if barely audible and not heard/present or not recorded in all auscultation locations; \\
 Grade II & if soft but easily heard in all auscultation locations; \\
 Grade III &  if moderately loud or loud. \\
\end{tabular}
\label{GradingTable}
\end{table}

Accordingly, the grade annotations can diverge from the original definition of murmur grading, when applied to cases for which not all the auscultation locations are available. In such cases, murmurs were classified by default as grade I/VI. Moreover, the cases classified as grade III/VI, actually include murmurs that could potentially be of grade III/VI or higher, since discrimination among grades III/VI, IV/VI, V/VI, and VI/VI are associated with palpable murmurs, also known as a thrills~\cite{Levine:1933}, which can only be assessed via physical in-person examination.
A cardiopulmonologist manually classified and characterized murmur events blindly, and independent of other clinical notes.

The sounds were recorded in an ambulatory environment. Different noisy sources have been observed in our dataset, from the stethoscope rubbing noise to a crying or laughing sound in the background. Thus, the automatic analysis of CirCor DigiScope dataset is indeed a hard task. On the other hand, the proposed dataset is a representative sample of real case environments where CAD systems must operate. 

\section{Discussion}
\label{par:discussion}
Given the large number of participants from the CC2014 and CC2015 screening campaigns, the CirCor DigiScope dataset is a representative sample of the rural and urban populations in Northeast Brazil, Paraíba. The dataset focuses on a pediatric population, $63 \%$ of the patients were children, $20\%$ of the patients were infants. Furthermore, and due to the challenges associated with the care of individuals with complex CHD, exceptionally, a few young adults that voluntarily asked to participate in the screening campaigns were also examined.

Overall, the screened population demonstrates a generally good clinical condition, with most clinical and physiological parameters within the normal range for the age. 
Despite the generally good clinical condition of the participants, a variety of congenital and acquired diseases have been found. Furthermore, not only simple cardiopathies, but also severe and complex cardiopathies that required specialist referral for advanced treatment were also found. In contrast with CC2014, in the CC2015 campaign, most participants were referred for follow-up (61.6\%), while 32.5\% were discharged. Seven patients (0.8\%) had a severe pathology with, requiring surgery or intervention. Further details are provided in Table~\ref{IndicationDiagnosisPlanning2014and2015}. 

The high number of pathologies encountered within the screened population also translated into a significant number of patients presenting murmurs in their auscultations. 
Most of the murmurs were observed in the systolic period ($96.8\%$), and only a few cases were reported in the diastolic period ($3.2\%$), see Table \ref{CC2014_2015Indicators}. This is due to the fact that the pressure gradient felt in the aortic and pulmonary valves, during the systolic ejection is very high when compared to the pressure gradient felt in the Mitral and Tricuspid valve, during the ventricular relaxation \cite{Douglas:2015}. As a result, stenosis in the Aortic and Pulmonary valves are more often than in the Mitral and Tricuspid valves \cite{Oktay2017}, \cite{Kim2019}, \cite{SNELLEN1968}. 
This is consistent with the observation that most of the complex congenital cardiopathies are observed in the systolic period \cite{Brunetti2015}. Furthermore, innocent murmurs, which are very common in infant populations, are mostly observed at the beginning of the systolic period \cite{Brunetti2015}.  Finally, from the technical perspective the auscultation of diastolic murmurs is a difficult procedure, since these murmurs are faint and harder to detect \cite{Butterworths:1990}. For example, in the recordings of the patient with the identifier \texttt{49824} and \texttt{66421}, early and middle diastolic murmurs were detected, respectively.   

Regarding the murmur timing analysis, the most common murmur is Holosystolic, which is a murmur that lasts over the full systolic period, see Table \ref{CC2014_2015Indicators}. This kind of murmur is commonly observed in ventricular septal defect pathologies \cite{Ampie2015}.
Holosystolic murmur waves were detected, for example in the recordings of the patient with the identifier \texttt{49628}. Early-systolic murmurs are also common in children, these murmurs happen immediately at the beginning of each systolic period and disappear shortly before the mid-systolic period. Usually such murmurs are innocent  \cite{Doshi2018}. Early-systolic murmur waves were observed, for example in the recordings of the patient with the identifier \texttt{49691}. Note that in the provided dataset, the murmur waves can be easily extracted by using the segmentation data (S1 and S2 wave locations) provided for each heart beat.

The majority of the detected murmurs have a Plateau shape, i.e., their intensities are approximately constant over time. Murmurs with such shapes can be associated to ventricular septal defect, a very common pathology in our database \cite{Ampie2015}. Murmur waves with a Plateau shape were detected, for example in the recordings of the patient with the identifier \texttt{50159}.
Note that the diamond-shaped murmurs (18.5\%) are usually more prevalent within a pediatric population than murmurs with plateau shape (59.8\%) \cite{LIEBMAN1968}. The large amount of plateau murmurs observed in the dataset might be explained by the fact that digital filters can attenuate or modify the murmur shape.  Murmur waves with a Diamond shape were observed, for example in the recordings of the patient with the identifier \texttt{50724}.

Some innocent murmurs are ``musical'' in quality (1.3\%, which is extremely a rare event) \cite{Stein1983}. A ``harsh'' murmur is described by a high-velocity blood flow from a higher to a lower pressure gradient \cite{Amnon:1984}. The term ``harsh'' is appropriate for describing the murmur in patients with significant semilunar valve stenosis or a ventricular septal defect (52.5\%). Murmur waves with a Harsh quality were detected, for example in the recordings of the patient with the identifier \texttt{49628}.
A ``blowing'' murmur is a sound caused by turbulent (rough) blood flow through the heart valves. In the Apex, it might be related to a mitral valve regurgitation (46.2\%)~\cite{Amnon:1984}.Murmur waves with a Blowing quality were observed, for example in the recordings of the patient with the identifier \texttt{49754}.

The variable ``grade'' refers to intensity, i.e., the loudness of the murmur. The location of the lesion and the distance to the stethoscope affects the listener's perception. Usually, louder murmurs (grade $\geq$ 3) are more likely to represent cardiac defects than silent ones \cite{Keren:2005}. Regarding systolic murmurs, the Levine scale is a numeric score used to characterize the intensity or the loudness of a murmur \cite{Levine:1933}. In the presented dataset, 27\% of the systolic murmurs and 10\% of the diastolic murmurs were grade III or greater, see Table \ref{CC2014_2015Indicators}. Murmur waves greater or equal than grade III were detected, for example in the patient with the identifier \texttt{50308}.

The analysis of the pitch is another important variable. The higher the pitch, the higher is the pressure gradient between the heart chambers. For example, aortic stenosis has a higher pitched murmur than a mitral stenosis. Furthermore, and in agreement with past observations, murmurs generated from a ventricular septal defect have usually a low pitch \cite{Ampie2015}. We should note however that pitch analysis from auscultation is not trivial or easy to describe; since filtering effects on a listener's perception are not completely characterizable. The pitch quality also varies across different stethoscopes. This is technically due to the difference between the transfer functions of different stethoscopes (and the preprocessing filters, in digital stethoscope front-end). This issue can be partially mitigated in digital auscultation, by designing \textit{digital equalizers} that compensate for the gain losses and amplitude/phase distortions of the analog front-end. Equalizers can also be applied to make the auscultations sound more similar to their analog counterparts, which are more familiar for expert annotators. The theory and practice of sound equalization have been extensively studied in the context of audio signal processing \cite{zolzer2008digital,bharitkar2008immersive}. Low, medium and high pitch murmur waves were detected, for example in the recordings of the patient with the identifier \texttt{49693}, \texttt{49825} and \texttt{49712}, respectively.

The location where the systolic, diastolic or systolic-diastolic murmurs were detected with the highest intensity is also an important feature to analyze. A damaged valve usually generates a murmur louder in its corresponding auscultation area. This is mainly due to the proximity and vicinity between the auscultation point and the damaged valve. For example, a murmur caused by aortic stenosis is often best heard at the upper sternal border, and usually on the right side \cite{Thomas:2020}. In this dataset, most of the murmurs are detected with a highest intensity in the pulmonary point, see Table \ref{CC2014_2015Indicators}. Note that in infants and children (up to three years old), the pulmonary and tricuspid spots partially overlap and, in most cases, a single recording auscultation location is used instead \cite{Carlo2014}. Consequently, heart sounds collected from these locations in children can be very similar to each other. Therefore it is not surprising that the pulmonary and tricuspid are the best auscultation locations to detect cardiac murmurs in our dataset.   For example, the patient with the identifier \texttt{50052}, murmur waves were detected more intensively in the aortic spot.


The observed murmurs do radiate and spread homogeneously through the thorax (cf. Table \ref{CC2014_2015Indicators}). Thus easily detected (almost equally) in many auscultation locations.


As a final remark, the CirCor DigiScope dataset is by far the largest publicly available heart sound dataset (5282 recordings), including recordings collected from multiple auscultation locations on the body. Moreover, the majority of the heart sounds (215,780 in total) were manually segmented and their quality was assessed by two independent cardiac physiologists. Furthermore,  the study resulted in a very detailed murmur characterization and classification database (including timing, pitch, grading, shape, quality, auscultation location, etc.), which can be used in future research.

\section{Conclusions}
\label{par:conclusions}
The CirCor DigiScope dataset represents a unique cohort from a pediatric population,  and a pregnant population, with significant congenital and acquired cardiac diseases. The age distribution is also homogeneously distributed in our dataset, thus potentially paving the way to the design of robust decision support systems for different target populations, from neonates to adults.

Given the rich annotations and characterizations provided with CirCor DigiScope dataset, the current work can be leveraged in various ways, including the design of multi-channel systems for multi-site PCG analysis, detection and classification of heart murmurs; and the automatic generation of murmur reports. For future work, we intend to make a comparative study, concerning the segmentation and classification of heart sound signals, in the CirCor DigiScope dataset. In future screening campaigns a broader population from different age groups shall be screened for CHD. Furthermore, new data collection protocols and technologies are going to be proposed and developed, aiming to collect data from the four auscultation spot in a synchronous or asynchronous manner.  

\section{Acknowledgment}
This work is a result of the Project DigiScope2 (POCI-01-0145-FEDER-029200-PTDC/CCI-COM/29200/2017) funded by Fundo Europeu de Desenvolvimento Regional (FEDER), through Programa Operacional Competitividade e Internacionalização (POCI), and by national funds, through Fundação para a Ciência e Tecnologia (FCT). This work is also financed by National Funds through the Portuguese funding agency, FCT - Fundação para a Ciência e a Tecnologia, within project UIDB/50014/2020. F.~Renna also acknowledges national funds through FCT -- Funda\c{c}\~{a}o para a Ci\^{e}ncia e a Tecnologia, I.P., under the Scientific Employment Stimulus-- Individual Call-- CEECIND/01970/2017. A.~Elola receives financial support by the Spanish Ministerio de Ciencia, Innovación y Universidades through grant RTI2018-101475-BI00, jointly with the Fondo Europeo de Desarrollo Regional (FEDER), and by the Basque Government through grant IT1229-19. 
G.~D.~Clifford is partly funded by 
the National Institute of Biomedical Imaging and Bioengineering (NIBIB) under grant \# R01EB030362.

The authors would like to acknowledge the support and efforts of the 2014 and 2015 Caravana do Coração screening campaign members. 

\bibliographystyle{IEEEtran}

\bibliography{references}

\begin{thebibliography}{10}
\providecommand{\url}[1]{#1}
\csname url@samestyle\endcsname
\providecommand{\newblock}{\relax}
\providecommand{\bibinfo}[2]{#2}
\providecommand{\BIBentrySTDinterwordspacing}{\spaceskip=0pt\relax}
\providecommand{\BIBentryALTinterwordstretchfactor}{4}
\providecommand{\BIBentryALTinterwordspacing}{\spaceskip=\fontdimen2\font plus
\BIBentryALTinterwordstretchfactor\fontdimen3\font minus
  \fontdimen4\font\relax}
\providecommand{\BIBforeignlanguage}[2]{{%
\expandafter\ifx\csname l@#1\endcsname\relax
\typeout{** WARNING: IEEEtran.bst: No hyphenation pattern has been}%
\typeout{** loaded for the language `#1'. Using the pattern for}%
\typeout{** the default language instead.}%
\else
\language=\csname l@#1\endcsname
\fi
#2}}
\providecommand{\BIBdecl}{\relax}
\BIBdecl

\bibitem{Douglas:2015}
P.~Libby, R.~Bonow, D.~Mann, and D.~Zipes, \emph{\BIBforeignlanguage{English
  (US)}{Braunwald's Heart Disease: A Textbook of Cardiovascular Medicine. 8th
  edition}}.\hskip 1em plus 0.5em minus 0.4em\relax Elsevier Science, 2007.

\bibitem{WHO:2017}
``{World Health Organization} -- cardiovascular diseases {(CVDs)},'' [Online].
  Available: \url{http://www.who.int/mediacentre/ factsheets/fs317/en/},
  accessed: 2021-05-10.

\bibitem{Gheorghe:2018}
A.~Gheorghe, U.~Griffiths, A.~Murphy, H.~Legido-Quigley \emph{et~al.}, ``The
  economic burden of cardiovascular disease and hypertension in low- and
  middle-income countries: {A} systematic review,'' \emph{BMC Public Health},
  vol.~18, 12 2018.

\bibitem{Murphy:2020}
A.~Murphy, B.~Palafox, M.~Walli-Attaei, T.~Powell-Jackson \emph{et~al.}, ``The
  household economic burden of non-communicable diseases in 18 countries,''
  \emph{BMJ Global Health}, vol.~5, no.~2, 2020.

\bibitem{Benjamin:2019}
\BIBentryALTinterwordspacing
E.~J. Benjamin, P.~Muntner, A.~Alonso, M.~S. Bittencourt \emph{et~al.}, ``Heart
  disease and stroke statistics{\textemdash}2019 update: A report from the
  american heart association,'' \emph{Circulation}, vol. 139, no.~10, pp.
  e56--e528, Mar. 2019. [Online]. Available:
  \url{https://doi.org/10.1161/cir.0000000000000659}
\BIBentrySTDinterwordspacing

\bibitem{Wilkins:2017}
E.~Wilkins, L.~Wilson, K.~Wickramasinghe, P.~Bhatnagar \emph{et~al.},
  \emph{\BIBforeignlanguage{English}{European Cardiovascular Disease Statistics
  2017}}.\hskip 1em plus 0.5em minus 0.4em\relax Belgium: European Heart
  Network, 2 2017.

\bibitem{Leal:2006}
\BIBentryALTinterwordspacing
J.~Leal, R.~Luengo-Fernández, A.~Gray, S.~Petersen, and M.~Rayner, ``{Economic
  burden of cardiovascular diseases in the enlarged European Union},''
  \emph{European Heart Journal}, vol.~27, no.~13, pp. 1610--1619, 02 2006.
  [Online]. Available: \url{https://doi.org/10.1093/eurheartj/ehi733}
\BIBentrySTDinterwordspacing

\bibitem{Soler:2000}
\BIBentryALTinterwordspacing
J.~Soler-Soler and E.~Galve, ``Worldwide perspective of valve disease,''
  \emph{Heart}, vol.~83, no.~6, pp. 721--725, 2000. [Online]. Available:
  \url{https://heart.bmj.com/content/83/6/721}
\BIBentrySTDinterwordspacing

\bibitem{Wealth:2011}
S.~Mendis, P.~Puska, B.~Norrving, W.~H. Organization \emph{et~al.},
  \emph{Global atlas on cardiovascular disease prevention and control}.\hskip
  1em plus 0.5em minus 0.4em\relax World Health Organization, 2011.

\bibitem{MANGIONE_2001}
\BIBentryALTinterwordspacing
S.~Mangione, ``{Cardiac auscultatory skills of physicians-in-training: a
  comparison of three English-speaking countries},'' \emph{The American Journal
  of Medicine}, vol. 110, no.~3, pp. 210 -- 216, 2001. [Online]. Available:
  \url{http://www.sciencedirect.com/science/article/pii/S0002934300006732}
\BIBentrySTDinterwordspacing

\bibitem{Narula:2018}
J.~Narula, Y.~Chandrashekhar, and E.~Braunwald, ``{Time to add a fifth pillar
  to bedside physical examination inspection, palpation, percussion,
  auscultation, and insonation},'' \emph{Journal of the American Medical
  Association Cardiology}, vol.~3, no.~4, pp. 346--350, 02 2018.

\bibitem{Falleni:2017}
S.~{Falleni}, A.~{Filippeschi}, E.~{Ruffaldi}, and C.~A. {Avizzano},
  ``Teleoperated multimodal robotic interface for telemedicine: A case study on
  remote auscultation,'' in \emph{2017 26th IEEE International Symposium on
  Robot and Human Interactive Communication (RO-MAN)}, Aug 2017, pp. 476--482.

\bibitem{Clifford:2016}
G.~D. {Clifford}, C.~{Liu}, B.~{Moody}, D.~{Springer} \emph{et~al.},
  ``{Classification of normal/abnormal heart sound recordings: The
  PhysioNet/Computing in Cardiology Challenge 2016},'' in \emph{2016 Computing
  in Cardiology Conference (CinC)}, Sep. 2016, pp. 609--612.

\bibitem{Pascal:2013}
E.~Gomes, P.~Bentley, M.~Coimbra, E.~Pereira, and Y.~Deng, ``{Classifying heart
  sounds: Approaches to the PASCAL challenge},'' \emph{HEALTHINF 2013 -
  Proceedings of the International Conference on Health Informatics}, pp.
  337--340, 01 2013.

\bibitem{Xiao:2019}
B.~{Xiao}, Y.~{Xu}, X.~{Bi}, W.~{Li} \emph{et~al.}, ``{Follow the Sound of
  Children’s Heart: A Deep Learning-based Computer-aided Pediatric CHDs
  Diagnosis System},'' \emph{IEEE Internet of Things Journal}, pp. 1--1, 2019.

\bibitem{Mattos:2015}
S.~d.~S. Mattos, S.~M.~V. Hazin, C.~T. Regis, J.~S. S.~d. Ara{\'u}jo
  \emph{et~al.}, ``{A telemedicine network for remote paediatric cardiology
  services in North-East Brazil},'' \emph{Bulletin of the World Health
  Organization}, vol.~93, pp. 881--887, 2015.

\bibitem{Mattos:2018}
S.~d.~S. Mattos, F.~A. Mourato, J.~S.~S. de~Ara{\'u}jo, L.~R. D.~N. Moser
  \emph{et~al.}, ``{Impact of a telemedicine network on neonatal mortality in a
  state in Northeast Brazil},'' \emph{Population health management}, vol.~21,
  no.~6, pp. 517--517, 2018.

\bibitem{Oliveira:2022}
J.~Oliveira, F.~Renna, P.~Costa, M.~Nogueira \emph{et~al.}, ``{The CirCor
  DigiScope Phonocardiogram Dataset (version 1.0.0)}, doi =
  {https://doi.org/10.13026/g02k-a047}.''

\bibitem{Dornbush:2020}
\BIBentryALTinterwordspacing
T.~A. Dornbush~S. (2019) Physiology, heart sounds. [Online]. Available:
  \url{In: StatPearls [Internet]. Treasure Island (FL): StatPearls Publishing,
  https://www.ncbi.nlm.nih.gov/books/NBK541010/}
\BIBentrySTDinterwordspacing

\bibitem{Karnath:2002}
B.~Karnath and W.~Thornton, ``Auscultation of the heart,'' \emph{Hospital
  Physician}, vol.~38, no.~9, pp. 39--43, 09 2002.

\bibitem{Chizner:2008}
M.~Chizner, ``Cardiac auscultation: Rediscovering the lost art,'' \emph{Current
  problems in cardiology}, vol.~33, pp. 326--408, 08 2008.

\bibitem{PRAKASH:1978}
\BIBentryALTinterwordspacing
R.~Prakash, ``Second heart sound: A phono-echocardiographic correlation in 20
  cardiac patients,'' \emph{Journal of the American Geriatrics Society},
  vol.~26, no.~8, pp. 372--374, 1978. [Online]. Available:
  \url{https://onlinelibrary.wiley.com/doi/abs/10.1111/j.1532-5415.1978.tb03687.x}
\BIBentrySTDinterwordspacing

\bibitem{Thomas:2020}
\BIBentryALTinterwordspacing
J.~D. Pollock and A.~N. Makaryus, \emph{Physiology, Cardiac Cycle}.\hskip 1em
  plus 0.5em minus 0.4em\relax StatPearls Publishing, Treasure Island (FL),
  2020, {PMID: 29083687}. [Online]. Available:
  \url{http://europepmc.org/books/NBK459327}
\BIBentrySTDinterwordspacing

\bibitem{Hoeting:2017}
\BIBentryALTinterwordspacing
N.~M. Hoeting, C.~E. McCracken, M.~McConnell, D.~Sallee \emph{et~al.},
  ``Systolic ejection click versus split first heart sound: Are our ears
  deceiving us?'' \emph{Congenital Heart Disease}, vol.~12, no.~4, pp.
  417--420, 2017. [Online]. Available:
  \url{https://onlinelibrary.wiley.com/doi/abs/10.1111/chd.12460}
\BIBentrySTDinterwordspacing

\bibitem{Riknagel:2017}
\BIBentryALTinterwordspacing
D.~Riknagel, H.~Zimmermann, R.~Farlie, D.~Hammershøi \emph{et~al.},
  ``Separation and characterization of maternal cardiac and vascular sounds in
  the third trimester of pregnancy,'' \emph{International Journal of Gynecology
  \& Obstetrics}, vol. 137, no.~3, pp. 253--259, 2017. [Online]. Available:
  \url{https://obgyn.onlinelibrary.wiley.com/doi/abs/10.1002/ijgo.12151}
\BIBentrySTDinterwordspacing

\bibitem{watkinson2001art}
J.~Watkinson, \emph{The art of digital audio}, 3rd~ed.\hskip 1em plus 0.5em
  minus 0.4em\relax Focal Press, 2001.

\bibitem{oppenheim1999discrete}
A.~Oppenheim, R.~Schafer, and J.~Buck, \emph{{Discrete-time signal
  processing}}, ser. Prentice-Hall signal processing series.\hskip 1em plus
  0.5em minus 0.4em\relax Prentice Hall, 1999.

\bibitem{Physionet2}
C.~Liu, D.~Springer, Q.~Li, and et. al., ``An open access database for the
  evaluation of heart sound algorithms,'' \emph{Physiological Measurement},
  vol.~37, no.~12, p. 2181, 2016.

\bibitem{Spadaccini:2013}
A.~{Spadaccini} and F.~{Beritelli}, ``Performance evaluation of heart sounds
  biometric systems on an open dataset,'' in \emph{2013 18th International
  Conference on Digital Signal Processing (DSP)}, July 2013, pp. 1--5.

\bibitem{Oliveira18st}
J.~Oliveira, F.~Renna, and M.~T. Coimbra, ``Adaptive sojourn time {HSMM} for
  heart sound segmentation,'' \emph{IEEE J. Biomed. Health Informatics},
  vol.~23, no.~2, pp. 642--649, 2019.

\bibitem{Cesarelli:2012}
\BIBentryALTinterwordspacing
M.~Cesarelli, M.~Ruffo, M.~Romano, and P.~Bifulco, ``Simulation of foetal
  phonocardiographic recordings for testing of fhr extraction algorithms,''
  \emph{Computer Methods and Programs in Biomedicine}, vol. 107, no.~3, pp. 513
  -- 523, 2012. [Online]. Available:
  \url{http://www.sciencedirect.com/science/article/pii/S0169260711003130}
\BIBentrySTDinterwordspacing

\bibitem{EPHNOGRAMDataset}
\BIBentryALTinterwordspacing
A.~Kazemnejad, P.~Gordany, and R.~Sameni, ``{EPHNOGRAM: A Simultaneous
  Electrocardiogram and Phonocardiogram Database},'' 2021. [Online]. Available:
  \url{https://physionet.org/content/ephnogram/1.0.0/}
\BIBentrySTDinterwordspacing

\bibitem{Springer:2016}
D.~B. Springer, L.~Tarassenko, and G.~D. Clifford, ``Logistic
  regression-{HSMM}-based heart sound segmentation.'' \emph{IEEE Transactions
  on Biomedical Engineering}, vol.~63, no.~4, pp. 822--832, 2016.

\bibitem{Pereira:2011}
D.~Pereira, F.~Hedayioglu, R.~Correia, T.~Silva \emph{et~al.}, ``{DigiScope} -
  {Unobtrusive} collection and annotating of auscultations in real hospital
  environments,'' in \emph{Engineering in Medicine and Biology Society, IEEE
  Conference}, 2011, pp. 1193--1196.

\bibitem{KazemnejadGordanySameni2021}
\BIBentryALTinterwordspacing
A.~Kazemnejad, P.~Gordany, and R.~Sameni, ``An open-access simultaneous
  electrocardiogram and phonocardiogram database,'' \emph{bioRxiv}, 2021.
  [Online]. Available: \url{https://doi.org/10.1101/2021.05.17.444563}
\BIBentrySTDinterwordspacing

\bibitem{Williams:2012}
K.~Williams, D.~Thomson, I.~Seto, D.~Contopoulos-Ioannidis \emph{et~al.},
  ``Standard 6: Age groups for pediatric trials,'' \emph{Pediatrics}, vol. 129
  Suppl 3, pp. S153--60, 06 2012.

\bibitem{Gomes:2016}
P.~Gomes, S.~Faria, and M.~Coimbra, ``The effect of data exchange protocols on
  decision support systems for heart sounds,'' vol. 2016, 08 2016, pp.
  5384--5387.

\bibitem{Renna19}
F.~{Renna}, J.~H. {Oliveira}, and M.~T. {Coimbra}, ``Deep convolutional neural
  networks for heart sound segmentation,'' \emph{IEEE Journal of Biomedical and
  Health Informatics}, vol.~23, no.~6, pp. 2435--2445, 2019.

\bibitem{Owen2015}
\BIBentryALTinterwordspacing
S.~J. Owen and K.~Wong, ``\BIBforeignlanguage{eng}{Cardiac auscultation via
  simulation: a survey of the approach of uk medical schools},''
  \emph{\BIBforeignlanguage{eng}{BMC research notes}}, vol.~8, pp. 427--427,
  Sep 2015, 26358413[pmid]. [Online]. Available:
  \url{https://pubmed.ncbi.nlm.nih.gov/26358413}
\BIBentrySTDinterwordspacing

\bibitem{Levine:1933}
A.~Freeman and S.~Levine, ``The clinical significance of the systolic murmur. a
  study of 1000 consecutive “non-cardiac” cases,'' \emph{Ann Intern Med},
  vol.~6, p. 1371–1385, 1933.

\bibitem{Oktay2017}
\BIBentryALTinterwordspacing
A.~A. Oktay, Y.~E. Gilliland, C.~J. Lavie, S.~J. Ramee \emph{et~al.},
  ``Echocardiographic assessment of degenerative mitral stenosis: A diagnostic
  challenge of an emerging cardiac disease,'' \emph{Current Problems in
  Cardiology}, vol.~42, no.~3, pp. 71--100, Mar. 2017. [Online]. Available:
  \url{https://doi.org/10.1016/j.cpcardiol.2017.01.002}
\BIBentrySTDinterwordspacing

\bibitem{Kim2019}
\BIBentryALTinterwordspacing
M.~S. Kim, S.~J. Cho, S.-J. Park, S.~W. Cho \emph{et~al.}, ``Frequency and
  clinical associating factors of valvular heart disease in asymptomatic korean
  adults,'' \emph{Scientific Reports}, vol.~9, no.~1, Nov. 2019. [Online].
  Available: \url{https://doi.org/10.1038/s41598-019-53277-0}
\BIBentrySTDinterwordspacing

\bibitem{SNELLEN1968}
\BIBentryALTinterwordspacing
H.~A. Snellen, H.~Hartman, T.~N. Buis-Liem, E.~H. Kole, and J.~Rohmer,
  ``Pulmonic stenosis,'' \emph{Circulation}, vol.~38, no. 1s5, Jul. 1968.
  [Online]. Available: \url{https://doi.org/10.1161/01.cir.38.1s5.v-93}
\BIBentrySTDinterwordspacing

\bibitem{Brunetti2015}
\BIBentryALTinterwordspacing
N.~D. Brunetti, S.~Rosania, C.~D'Antuono, A.~D'Antuono \emph{et~al.},
  ``Diagnostic accuracy of heart murmur in newborns with suspected congenital
  heart disease,'' \emph{Journal of Cardiovascular Medicine}, vol.~16, no.~8,
  pp. 556--561, Aug. 2015. [Online]. Available:
  \url{https://doi.org/10.2459/jcm.0b013e3283649953}
\BIBentrySTDinterwordspacing

\bibitem{Butterworths:1990}
\BIBentryALTinterwordspacing
``Clinical methods: The history, physical, and laboratory examinations,'' 1990.
  [Online]. Available: \url{https://www.ncbi.nlm.nih.gov/books/NBK201}
\BIBentrySTDinterwordspacing

\bibitem{Ampie2015}
\BIBentryALTinterwordspacing
L.~E. Ampie and S.~EL-Amin, \emph{Ventricular Septal Defect}.\hskip 1em plus
  0.5em minus 0.4em\relax London: Springer London, 2015, pp. 289--297.
  [Online]. Available: \url{https://doi.org/10.1007/978-1-4471-6738-9_25}
\BIBentrySTDinterwordspacing

\bibitem{Doshi2018}
\BIBentryALTinterwordspacing
A.~R. Doshi, ``Innocent heart murmur,'' \emph{Cureus}, Dec. 2018. [Online].
  Available: \url{https://doi.org/10.7759/cureus.3689}
\BIBentrySTDinterwordspacing

\bibitem{LIEBMAN1968}
\BIBentryALTinterwordspacing
J.~Liebman and S.~Sood, ``Diastolic murmurs in apparently normal children,''
  \emph{Circulation}, vol.~38, no.~4, pp. 755--762, Oct. 1968. [Online].
  Available: \url{https://doi.org/10.1161/01.cir.38.4.755}
\BIBentrySTDinterwordspacing

\bibitem{Stein1983}
\BIBentryALTinterwordspacing
P.~D. Stein, H.~N. Sabbah, and J.~B. Lakier, ``Origin and clinical relevance of
  musical murmurs,'' \emph{International Journal of Cardiology}, vol.~4, no.~1,
  pp. 103--112, Aug. 1983. [Online]. Available:
  \url{https://doi.org/10.1016/0167-5273(83)90223-1}
\BIBentrySTDinterwordspacing

\bibitem{Amnon:1984}
\BIBentryALTinterwordspacing
A.~Rosenthal, ``How to distinguish between innocent and pathologic murmurs in
  childhood,'' \emph{Pediatric Clinics of North America}, vol.~31, no.~6, pp.
  1229--1240, 1984, symposium on Pediatric Cardiology. [Online]. Available:
  \url{https://www.sciencedirect.com/science/article/pii/S0031395516347186}
\BIBentrySTDinterwordspacing

\bibitem{Keren:2005}
\BIBentryALTinterwordspacing
R.~Keren, M.~Tereschuk, and X.~Luan, ``{Evaluation of a Novel Method for
  Grading Heart Murmur Intensity},'' \emph{Archives of Pediatrics and
  Adolescent Medicine}, vol. 159, no.~4, pp. 329--334, 04 2005. [Online].
  Available: \url{https://doi.org/10.1001/archpedi.159.4.329}
\BIBentrySTDinterwordspacing

\bibitem{zolzer2008digital}
U.~Z{\"o}lzer, \emph{Digital audio signal processing}.\hskip 1em plus 0.5em
  minus 0.4em\relax John Wiley \& Sons, 2008.

\bibitem{bharitkar2008immersive}
S.~Bharitkar and C.~Kyriakakis, \emph{Immersive audio signal processing}.\hskip
  1em plus 0.5em minus 0.4em\relax Springer Science \& Business Media, 2008.

\bibitem{Carlo2014}
\BIBentryALTinterwordspacing
C.~Ratti, L.~Grassi, E.~D. Maria, L.~Bonetti \emph{et~al.}, ``L’auscultazione
  cardiaca nel bambino,'' \emph{Recenti Progressi in Medicina}, no.
  2014Dicembre, Dec. 2014. [Online]. Available:
  \url{https://doi.org/10.1701/1706.18620}
\BIBentrySTDinterwordspacing

\end{thebibliography}

\end{document}